\documentclass[journal]{IEEEtran}
\usepackage{latexsym,amssymb,amsmath,graphicx,epsfig,bbm,float,epsf,graphics}
\usepackage{tabularx}
\usepackage{url}
\usepackage{multicol}
\usepackage{amsmath}
\usepackage{subcaption}
\usepackage{relsize}
\usepackage{tcolorbox}
\usepackage{mathtools}
\usepackage{enumitem}
\usepackage{multirow}
\usepackage[font=small,labelfont=bf]{caption}
\usepackage[ruled,vlined]{algorithm2e}
\usepackage{tikz}
\usepackage{dsfont}
\usepackage{booktabs}
\usepackage{cite}

\renewcommand{\vec}[1]{\mbox{\boldmath${#1}$}}

\DeclareMathOperator*{\argmax}{arg\,max}

\begin{document}
%
\title{Blind Reconstruction of Multilayered Tissue Profiles with UWB Radar Under Bayesian Setting}

\author{Burak Cevat Civek and
        Emre Ertin
\thanks{B. C. Civek and E. Ertin are with the Department
of Electrical and Computer Engineering, The Ohio State University, Columbus, OH, 43210, USA. Contact e-mail: civek.1@osu.edu}}


\maketitle

\begin{abstract}
In this paper, we investigate the problem of inverse electromagnetic scattering to recover multilayer human tissue profiles using ultrawideband radar systems in Bayesian setting. We study the recovery problem in blind setting, in which we simultaneously estimate both the dielectric/geometric properties of the one-dimensional target tissue profile and the transmitted radar waveform. To perform Bayesian parameter estimation, we propose a hybrid and adaptive Markov Chain Monte Carlo method, which combines the Slice sampling and Hamiltonian Monte Carlo approaches. The introduced sampling mechanism also incorporates the Parallel Tempering approach to escape from the local optimal regions of the complex posterior distribution. We provide empirical support through various numerical simulations for the achieved enhanced sampling efficiency compared to conventional sampling schemes. To investigate the recovery performance, we work on synthetic measurements simulating actual radar returns from multilayer tissue profiles. We derive theoretical bounds for the best achievable estimation performance in terms of normalized root mean square error and provide a comparison with the performance of our estimator.
\end{abstract}


\setlength{\abovedisplayskip}{2pt}
\setlength{\belowdisplayskip}{2pt}
\setlength{\textfloatsep}{0pt}
\setlength{\dbltextfloatsep}{0pt}
\setlength{\abovecaptionskip}{1pt}

\section{Introduction}
\IEEEPARstart{R}{emote} sensing of human physiology is of growing importance in medical research for the diagnosis and treatment of chronic diseases \cite{PantelopoulosA,MajumderS}. Monitoring the alterations in internal tissue composition provides valuable information about the progression of life-threatening diseases, including but not limited to, brain tumor, pulmonary edema, and cardiac disorders \cite{ShyamalP}. However, traditional imaging modalities, such as Magnetic Resonance Imaging (MRI), Computed Tomography (CT), or Ultrasound, are not feasible for monitoring variations regularly, e.g., on a daily basis, due to their high cost and accessibility issues. Therefore, more efficient, low-cost, and possibly mobile sensing schemes are needed for frequent and long-term measurements on the human body.\par

Following the advancements in sensor technologies, reliable characterization of tissue profiles is becoming viable for both clinic and home environments at much lower costs with easy access \cite{GaoJ2}. Specifically, ultrawideband (UWB) radar sensors emitting electromagnetic (EM) waves, which can penetrate through most of the biological tissues including skin, fat, muscle, etc., provide a promising alternative to those conventional sensing modalities \cite{ChinC,ZetikR}. In principle, a UWB radar system transmits a short duration pulse and records the backscattered signal composed of reflections from the target object. In human body, each tissue exhibits distinct dielectric properties, i.e., permittivity and conductivity. This causes impedance mismatches at the interfaces and creates multiple reflection points for the impinging transmitted pulse. Therefore, a rich backscattered signal, which is strongly affected by the dielectric properties, is observed and can be processed to make inferences about the tissue composition underneath the skin. \par

The emergence of UWB radar as a medical sensing technology occurred when McEwan described the physical principle of the UWB system which was able to detect movements of the heart wall in the two patents awarded to him \cite{McEwanT1,McEwanT}. Since then, detecting vital signs of human body, such as respiration and heart rate, is one of the most widely studied problems in medical UWB sensing \cite{ZetikR,DiasD}. Many studies successfully recovered vital signs in a non-invasive manner due to the sensitivity of the backscattered signal to movements of the inner tissues, such as lungs or heart \cite{GaoJ,GaoJ3}. In this work, however, instead of measuring vital signs, we focus on extracting a complete reflectivity profile for sub-skin tissue composition in terms of the dielectric and geometric properties. Possible applications include detecting or monitoring the evolution of breast cancer, brain tumor, water retention in lungs, or pulmonary edema.\par

In general, the inference methods for detecting alterations in tissue compositions focus on the explicit recovery of the dielectric properties, such as permittivity and conductivity, as well as the geometrical properties, such as thickness, of the target tissues based on the backscattered measurement. In medical UWB sensing literature, a homogeneous multilayer planar model is a reasonable and widely studied model to describe the anatomical structure of the human body \cite{StaderiniE,VarottoG,CavagnaroM,KetataM}. One of the common techniques for inverse EM scattering problems targeting multilayer homogeneous mediums is the layer stripping, which is extensively studied in GPR systems using UWB pulses to evaluate the physical and geometric properties of the subsurface earth layers \cite{SaarenketoT,ALQadiI,LoizosA,LahouarS}. Layer stripping is a time domain approach that estimates the constitutive parameters of each layer in a sequential manner, i.e., at each iteration, the algorithm estimates the properties of the top-most layer and removes its effect from the backscattered signal, progressively reconstructing each layer until all layers are reconstructed. The estimation procedure is usually based on the amplitude and time-of-arrival of the echos reflected from the layer interfaces. Therefore, success of the technique is closely related to accurate estimation of reflected pulse amplitudes and corresponding time delays, which requires clearly separated echos in time domain \cite{LahouarS,AfricanoM}. Although this requirement is satisfied for many geophysical applications due to greater thicknesses of earth layers, such clear separation is usually not possible for human tissues. Moreover, typical layer stripping techniques assume the multiple reflections are negligible as in \cite{SaarenketoT,ALQadiI,LeeJ}, illustrating the validity of this assumption for geophysical applications such as road pavement evaluation and ice sheet reconstruction. However, multiple reflections have a dominating effect when the target medium is human body \cite{StaderiniE,CavagnaroM}. Recently, Caorsi \textit{et al.} \cite{CaorsiS}, proposed a comprehensive layer stripping technique which uses a binary decision tree approach \cite{CaorsiS2} to detect and remove the pulses caused by multiple reflections to eliminate ambiguities. The proposed technique successfully classifies each echo as a direct or multiple reflection in the case of well-separated pulses with loss-less mediums (zero conductivities), but the performance significantly degrades if overlaps exist or the mediums have non-zero conductivities. As a result, application of layer stripping is limited for medical UWB sensing due to overlapping pulses, multiple reflections, and non-negligible conductivity losses. \par 

An alternative to the time-domain layer stripping approach is the EM inversion, which constructs a least squares problem (usually in frequency domain) to minimize the mean squared error between the actual and reconstructed measurements. The reconstructed measurement is obtained through a problem specific forward model governing the EM wave propagation in layered media and antenna responses. The optimization is performed on the constitutive parameters, i.e., permittivity, conductivity and thickness, to find the set of parameters achieving the best fit to the actual measurement. In \cite{SpagnoliniU}, Spagnolini compared EM inversion with layer stripping and demonstrated its promising capabilities in radar inverse problems. Unlike layer stripping, which only concerns the time delay and amplitude information, EM inversion completely utilizes the underlying physical interactions in EM wave propagation. Therefore, it eliminates the need for the strong simplifying assumptions and facilitates successful recovery even for the cases where there exist overlapping pulses, multiple reflections and non-zero conductivities.\par

Even though EM inversion approach has extensive practical applications in GPR literature, its utilization for medical sensing problems has not yet been investigated. To eliminate this gap, in this work, we employ the EM inversion approach for estimating the parameters of multilayer targets composed of human tissues. We restrict the scope of this work to a one-dimensional setting in which plane waves propagate through non-dispersive homogeneous planar mediums. Although this is a simplified version of the reality, it provides useful insights to develop more sophisticated imaging systems.\par

The contributions of this work can be summarized as follows. Firstly, we pose the problem as a blind deconvolution problem and simultaneously estimate both the transmitted waveform and the reflectivity profile to achieve self-calibration. In practice, the waveform generated within the radar circuitry is distorted by the antenna transmitter/receiver responses, and hence, the actual transmitted waveform is unknown without an appropriate calibration process. Traditional approaches for UWB radar inverse problems, therefore, assume calibrated antenna responses. Secondly, we study the problem in Bayesian setting and present a comprehensive and efficient Markov Chain Monte Carlo (MCMC) method to estimate the marginal posterior densities of the unknowns. Unlike the widely employed deterministic least squares approach, this enables us to perform additional posterior analyses, from which quantitative uncertainty measures about the estimations can be obtained through credibility intervals. Finally, we derive theoretical bounds on the estimation of multilayer model parameters in blind setting, which signify the best achievable error performance of any estimator. We note that even though the presented MCMC methods are designed for one-dimensional wave propagation model, they can be extended to the three-dimensional scenario.\par
 
The paper is organized as follows. We first introduce the wave propagation and measurement models in Section \ref{problem_setting}, followed by the description of the problem formulation under Bayesian setting in Section \ref{sec_prob_desc}. Then, in Sections \ref{sec_prop_gibbs_sampler} and \ref{sec_prop_hybrid_sampler}, we present the proposed MCMC method for sampling from the highly complex posterior distribution. We validate the proposed sampling schemes and provide a comparison between the derived theoretical bounds and the performance of the proposed estimator in Section \ref{sec_simulations}. We finalize our discussion in Section \ref{sec_conclusion} with concluding remarks and possible future research directions.

\begin{figure}[t!]
    \centering
    \includegraphics[width=0.48\textwidth]{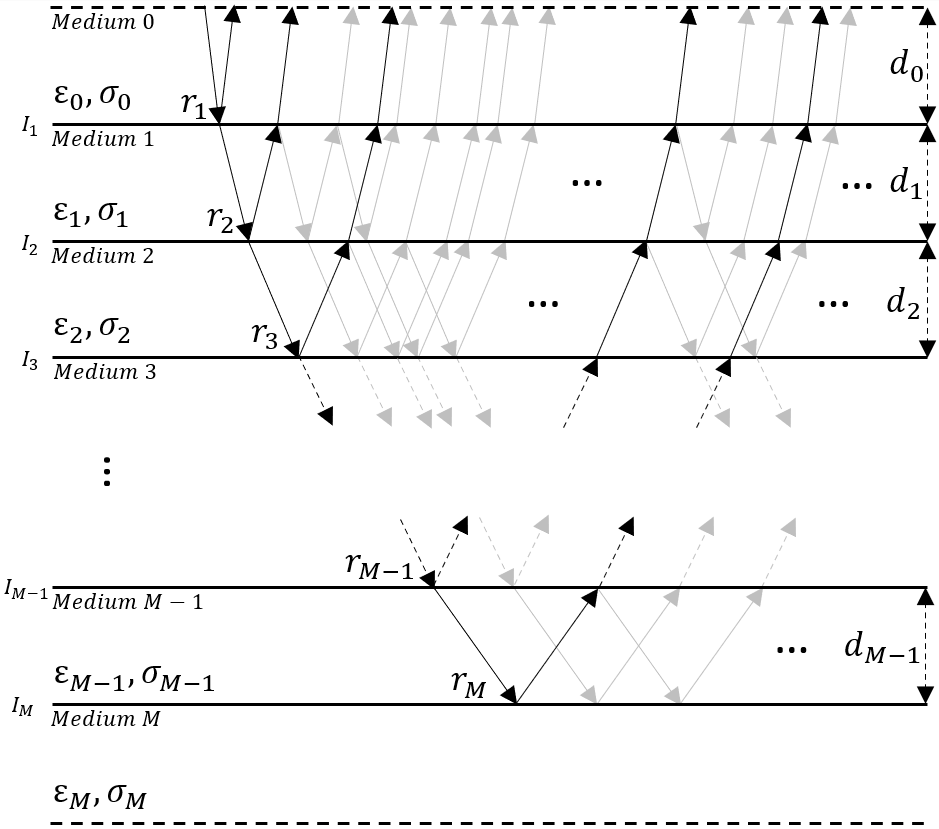}
    \caption{Illustration of reflection paths for an $M$-layer structure. Black arrows represent the primary reflection paths associated with each interface. Gray arrows represent the multiple bounces between the interfaces. Inclined arrows are used only for the illustration purposes.\label{fig_M_Layer_Model}}
\end{figure}
\vspace{-4mm}
\section{Measurement Model for Multilayer Reflectivity Profile}\label{problem_setting}

\subsection{Multilayer Reflection Model}
We consider an UWB system where we transmit a short duration UWB pulse and collect the backscattered signals which are reflections from an object composed of multiple planar layers. The layers are assumed to be homogeneous mediums and have distinct dielectric properties such that the interfaces between them can be considered as reflective surfaces. The backscattered signal can be expressed as a combination of scaled, shifted and distorted versions of the transmitted waveform. The distortion occurs due to materials either being dispersive or having non-zero conductivity. These factors are completely determined by the reflectivity profile of the target being monitored. In general, for an $M$-layer structure with thicknesses $d_i$, as illustrated in Fig. \ref{fig_M_Layer_Model}, where the last layer has infinite depth, the 1D downward reflectivity profile $X_i(\omega)$ in frequency domain has the following recursive form \cite{Chew}
\begin{equation}\label{multilayer_reflection_model}
    X_i(\omega) = \dfrac{r_i + X_{i+1}(\omega)e^{-2\alpha_i d_i}e^{-j2\beta_i d_i}}{1 + r_i X_{i+1}(\omega)e^{-2\alpha_i d_i}e^{-j2\beta_i d_i}},
\end{equation}
at each interface $I_i$ for $i = 1,\hdots,M-1$, with $X_M(\omega) = r_M$ and $\omega$ representing the angular frequency in rad/sec. The downward local reflection coefficient at interface $I_i$ is given by $r_i = (\eta_{i}-\eta_{i-1})/(\eta_{i}+\eta_{i-1})$, where $\eta_i = \sqrt{(j\omega\mu_o)/(\sigma_i + j\omega\varepsilon_o\varepsilon_i)}$ is the complex valued intrinsic impedance defined in terms of the dielectric constant $\varepsilon_i$ and conductivity $\sigma_i$ in S/m of the mediums. Here, $\mu_o$ and $\varepsilon_o$ are constants representing the vacuum permeability in H/m and vacuum permittivity in F/m respectively. Lastly, $\alpha_i = \omega[\mu_o\varepsilon_o\varepsilon_i(\zeta_i-1)/2]^{1/2}$ and $\beta_i = \omega[\mu_o\varepsilon_o\varepsilon_i(\zeta_i+1)/2]^{1/2}$ represent the attenuation coefficients and the phase constants respectively, where $\zeta_i = \sqrt{1 + (\sigma_i/\omega\varepsilon_o\varepsilon_i)^2}$. 
\vspace{-2mm}
\subsection{Measurement Model}
In this work, we consider the scenario in which the source of the transmitted pulse is $d_0$ meters away from the interface $I_1$ with normal incidence. Therefore, for a given frequency $\omega$, the corresponding frequency component of the transmitted pulse, $H(\omega)$, is multiplied by $X_0(\omega) = X_1(\omega)e^{-2\alpha_0 d_0}e^{-j2\beta_0 d_0}$, yielding the following backscattering model $Y(\omega) = H(\omega)X_0(\omega)$, where $Y(\omega)$ represents the frequency domain representation of the backscattered signal. In practice, we observe the measurement sampled at frequencies $\{\omega_n\}_{n=0}^{N-1}$, which can be modeled as 
\begin{equation}\label{measurement_model}
    \vec{y} = \text{diag}(\vec{F}_Q\vec{h})\vec{x} + \vec{v},
\end{equation}
where $\vec{y},\vec{x} \in \mathbbm{C}^{N}$ are defined as $\vec{y} = [Y(\omega_0),\hdots,Y(\omega_{N-1})]^T$ and $\vec{x} = [X_0(\omega_0),\hdots,X_0(\omega_{N-1})]^T$, and the transmitted waveform is modeled in time domain as $\vec{h} \in \mathbbm{R}^Q$ to limit its duration with $Q$ samples in time domain. The matrix $\vec{F}_Q \in \mathbbm{C}^{N \times Q}$ represents the appropriately selected partial DFT matrix. We model the measurement noise by including a complex valued additive noise term $\vec{v} \in \mathbbm{C}^{N}$.

\begin{figure}[t!]
    \centering
    \includegraphics[width=0.9\linewidth]{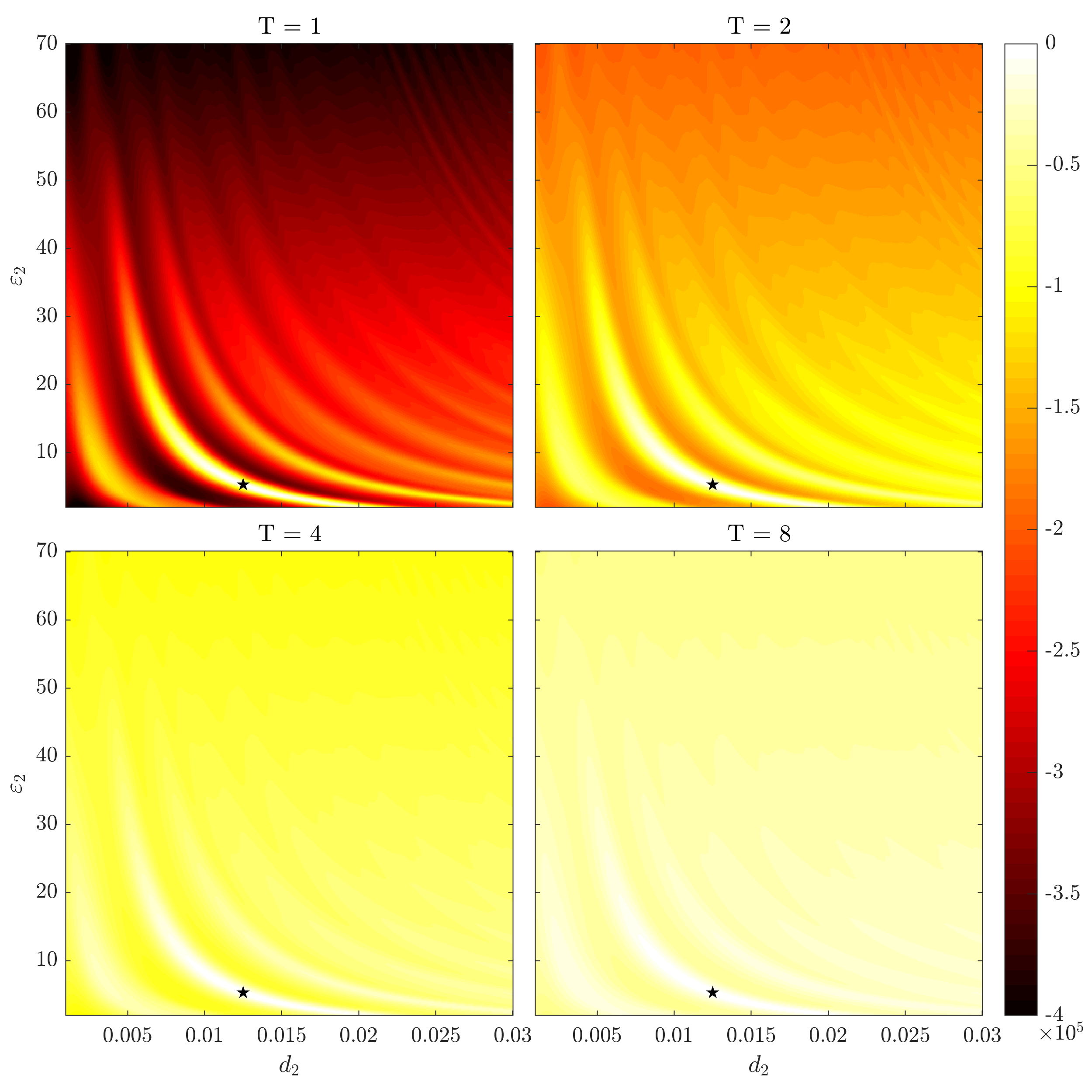}
    \caption{An example cross section of high dimensional log-posterior distribution $\log p(\vec{\theta},\vec{\gamma},\sigma_v^2|\vec{y})$ for $d_2$-$\varepsilon_2$ plane at different temperature levels. Remaining model parameters are fixed at their true values.\label{fig_posterior_2D_cross_sections}}
\end{figure}

\section{Problem Setting}\label{sec_prob_desc}
Our goal is to estimate the multilayer model parameters $\{\varepsilon_i\}_{i=1}^{M}$, $\{\sigma_i\}_{i=1}^{M}$, and $\{d_i\}_{i=0}^{M-1}$ along with the transmitted pulse $\vec{h}$ solely based on the measurement vector $\vec{y}$. We note that dielectric constant $\varepsilon_0$ (not to be confused with vacuum permittivity $\varepsilon_o$) and conductivity $\sigma_0$ of the first medium, where the source is located, are assumed to be known, but the distance $d_0$ between the transmitter and the first interface is also unknown and to be estimated. Following a Bayesian framework, we assign specific prior distributions on the unknown variables reflecting our prior knowledge, which are described in the subsequent sections. 

\subsubsection{Prior Distribution for Multilayer Model Parameters} 
We collect the multilayer model parameters in a single vector $\vec{\theta} = [\varepsilon_1,\hdots,\varepsilon_M,\sigma_1,\hdots,\sigma_{M},d_0,\hdots,d_{M-1}]^T$ for more compact notation. Assuming bounded parameter space $\Lambda_{\theta}$, where the lower and upper bounds are given by $\theta_{i,\text{min}}$ and $\theta_{i,\text{max}}$ for $i^{th}$ parameter, and statistically independent parameters, the joint prior distribution of $\vec{\theta}$ follows $p(\vec{\theta}) = \prod_{i=1}^{3M}p(\theta_i) =\prod_{i=1}^{3M}\mathcal{B}(\bar{\theta}_i;\lambda_i,\kappa_i)$
where $\mathcal{B}(\cdot;\lambda,\kappa)$ denotes the Beta distribution with mode $\lambda_i$, concentration $\kappa_i$, and $\bar{\theta}_{i} = (\theta_{i} - \theta_{i,\text{min}})/(\theta_{i,\text{max}} - \theta_{i,\text{min}})$. The individual parameters $\lambda_i$ and $\kappa_i$ are selected to reflect our prior knowledge.

\subsubsection{Prior Distribution for Pulse Sequence}
We represent the transmitted pulse $\vec{h} \in \mathbbm{R}^Q$ using a subspace $\vec{A} \in \mathbbm{R}^{Q \times L}$, i.e., $\vec{h} = \vec{A}\vec{\gamma}$, where $\vec{\gamma} \in \mathbbm{R}^L$ represents the random coefficient vector. Here, $\vec{A}$ is selected to reflect the frequency domain restrictions, i.e., it can be constructed by selecting the first $L$ sequence of either Discrete Prolate Spheroidal (DPS) Sequences or Hermite Functions \cite{FHlawatsch}. Instead of directly solving for $\vec{h}$, we solve for the coefficient vector $\vec{\gamma}$, which is assigned a zero-mean i.i.d. Gaussian distribution with known diagonal covariance $\vec{\Sigma}_\gamma = \text{diag}(\sigma_{\gamma}^2\vec{I})$, i.e., $p(\vec{\gamma}) = \mathcal{N}(\vec{\gamma};\vec{0},\vec{\Sigma}_{\gamma})$.

\subsubsection{Prior Distribution for Noise Variance} We model the measurement noise $\vec{v}$ with a circularly symmetric complex Gaussian law, $\mathcal{CN}(\vec{v};\vec{0},\sigma_v^2\vec{I})$, where its variance, $\sigma_v^2$, is another unknown and to be estimated along with the other model parameters. We assign Inverse-Gamma distribution with shape and scale parameters $\alpha_v$ and $\beta_v$ to noise variance since it is the analytically tractable conjugate prior for the unknown variance of Gaussian distribution, i.e., $p(\sigma_v^2) = \mathcal{IG}(\sigma_v^2;\alpha_v,\beta_v)$.\par 

Given the prior distributions for each of the variables, and assuming $\vec{\theta}$, $\vec{\gamma}$ and $\sigma_v^2$ are statistically independent, the posterior distribution has the following expression
\begin{equation}\label{posterior_dist}
    p(\vec{\theta},\vec{\gamma},\sigma_v^2|\vec{y}) \propto p(\vec{y}|\vec{\theta},\vec{\gamma},\sigma_v^2)p(\vec{\theta})p(\vec{\gamma})p(\sigma_v^2),
\end{equation}
where we dropped the irrelevant scaling factor $p(\vec{y})$. The likelihood term has the form of circularly symmetric complex Gaussian distribution
\begin{equation}\label{likelihood}
    p(\vec{y}|\vec{\theta},\vec{\gamma},\sigma_v^2) = \bigg(\dfrac{1}{\pi\sigma_v^2}\bigg)^{N}\exp\bigg(-\dfrac{\|\vec{y} - \text{diag}(\vec{B}\vec{\gamma})\vec{x}\|^2}{\sigma_v^2}\bigg)
\end{equation}
where $\vec{B} = \vec{F}_{Q}\vec{A}$ and $\|\cdot\|$ represents the $\ell_2$-norm of a vector.\par

We consider the Minimum Mean Square Error (MMSE) estimator, given by
\begin{equation}\label{MMSE_estimator}
    (\vec{\theta}^*,\vec{\gamma}^*,\sigma_v^{2*})_{\text{MMSE}} = E[\vec{\theta},\vec{\gamma},\sigma_v^2|\vec{y}],
\end{equation}
for the estimation of the parameters. However, the posterior distribution given in (\ref{posterior_dist}) is highly complex, possibly having multimodal structure with many local maxima, as illustrated in Fig. \ref{fig_posterior_2D_cross_sections}. In such cases, the Maximum A Posteriori (MAP) estimator could be a more favorable choice. Therefore, we also consider the MAP estimator, given by
\begin{equation}\label{MAP_estimator}
    (\vec{\theta}^*,\vec{\gamma}^*,\sigma_v^{2*})_{\text{MAP}} = \argmax_{\vec{\theta},\vec{\gamma},\sigma_v^2}p(\vec{\theta},\vec{\gamma},\sigma_v^2|\vec{y}).
\end{equation}
\par
The MMSE estimator requires intractable integration of the posterior distribution due to its complex structure. The MAP estimator, on the other hand, can be achieved by employing off-the-shelf gradient ascent methods, since the probability space is well-defined and does not have any discontinuities. However, due to existence of many local maxima, initialization plays a critical role on finding the global maximum. Therefore, we propose to employ MCMC simulations, which not only provide an approximate MMSE solution through the sample mean, but also explore the high probability regions of the parameter space, yielding a good initialization for achieving the MAP solution. Moreover, besides the point estimates, this approach also enables us to calculate credibility intervals to represent uncertainties about the estimations.

\section{Gibbs Sampler with Parallel Tempering}\label{sec_prop_gibbs_sampler}
The MCMC simulations are widely used in complex Bayesian inference problems to achieve numerical solutions. The core of the MCMC methods is the samplers, which are used to draw samples from a target distribution, which is the posterior distribution given in (\ref{posterior_dist}) in our case. These samples can then be used to approximate the statistics of the target distribution, for example, the MMSE estimation can be approximated by the mean average of the samples drawn from the posterior distribution. However, the multimodality of the posterior distribution significantly reduces the efficiency of the MCMC samplers, i.e., although the probability of jump from one mode to another is not zero, it is generally small enough, causing the sampler to get stuck on one mode of the distribution for a long time. In order to resolve this issue, we adopt a tempering approach, i.e., Parallel Tempering, which substantially improves the exploration power when combined with the standard MCMC samplers. In this section, we first briefly discuss the general idea of tempering and specifically the Parallel Tempering, followed by the description of our proposed MCMC sampler.

\subsection{Tempering Approaches for Multimodal Distributions}
Consider a high dimensional target probability distribution $\pi(\vec{z})$, from which we aim to draw samples. When the target distribution $\pi(\vec{z})$ is highly multimodal, the standard MCMC samplers such as MH and Gibbs, or even more sophisticated methods like HMC, fail to explore the probability space efficiently, due to the low probability regions acting like barriers in between the modes of the distribution. The main idea of tempering is to augment the original target distribution $\pi(\vec{z})$ with an additional temperature variable $T$ to create the tempered distribution $\pi(\vec{z};T) = K(T)\pi(\vec{z})^{1/T}$, where $K(T)$ denotes the normalization constant. As illustrated in Fig. \ref{fig_posterior_2D_cross_sections}, tempering, when $T > 1$, has a flattening effect on the original distribution, which removes the low probability barriers between the modes. Therefore, jumps between different modes become much more likely for the distributions with high temperatures.\par 

The idea of Parallel Tempering (PT) is to run multiple MCMC chains independently and simultaneously at each temperature level with stochastic temperature swaps between the neighbouring temperature levels \cite{Geyer}. The target distribution in PT is a joint distribution over all chains given by $\prod_{\ell=1}^{L} \pi(\vec{z}^{(\ell)};T_{\ell})$, where $\vec{z}^{(\ell)}$ denotes the variables for the chain running at temperature level $T_{\ell}$. Assuming symmetric proposals, the acceptance probability $\alpha_{\ell,\ell+1}$ that maintains the detailed balance in the case of a temperature swap between the chains at $T_{\ell}$ and $T_{\ell+1}$ is given by
\begin{equation}\label{MH_PT_acceptance}
    \alpha_{\ell,\ell+1} = \min\bigg\{1,\dfrac{\pi(\vec{z}^{(\ell)})^{1/T_{\ell+1}} \pi(\vec{z}^{(\ell+1)})^{1/T_{\ell}}}{\pi(\vec{z}^{(\ell+1)})^{1/T_{\ell+1}}\pi(\vec{z}^{(\ell)})^{1/T_{\ell}}}\bigg\}.
\end{equation}
\vspace{-5mm}
\subsection{Proposed Gibbs Sampler with Parallel Tempering}
We begin with introducing the general structure of our proposed sampler and discussing its connection to the Parallel Tempering approach. We employ a Gibbs sampler scheme, which is a powerful MCMC tool for sampling from high dimensional distributions especially when the conditional posteriors are analytically tractable and straightforward to sample from \cite{SGeman}. Here, note that the multimodality of the posterior is mainly due to the likelihood function given in (\ref{likelihood}). The prior distributions assigned to the pulse shape and the noise variance do not contribute to the multimodality of the target posterior. Therefore, we follow an alternative tempering approach, where we partially temper the posterior distribution by applying tempering only to the likelihood. With this approach, the chains running at high temperatures will sample from the prior distributions, instead of a flat distribution over the parameter space. This is quite useful when the prior distributions are unimodal, which is the case for the Gaussian and Inverse-Gamma distributions.\par 
\begin{table}[t]
\centering
\small
\caption{Proposed Gibbs sampler for partially tempered posterior distribution $p(\vec{\theta},\vec{\gamma},\sigma_v^2|\vec{y};T)$ for a given temperature $T$.}
\label{tab_proposed_gibbs_sampler}
\renewcommand\arraystretch{1.5}
\begin{tabular}{|m{0.461\textwidth}|}
\hline
Step 1. Draw $\sigma_v^2$ from $p(\sigma_v^2|\vec{y},\vec{\theta},\vec{\gamma};T) \propto p(\vec{y}|\vec{\theta},\vec{\gamma},\sigma_v^2)^{1/T}p(\sigma_v^2)$   \\
Step 2. Draw $\vec{\gamma}$ from $p(\vec{\gamma}|\vec{y},\vec{\theta},\sigma_v^2;T) \propto p(\vec{y}|\vec{\theta},\vec{\gamma},\sigma_v^2)^{1/T}p(\vec{\gamma})$   \\
Step 3. Draw $\vec{\theta}$ from $p(\vec{\theta}|\vec{y},\vec{\gamma},\sigma_v^2;T) \propto p(\vec{y}|\vec{\theta},\vec{\gamma},\sigma_v^2)^{1/T}p(\vec{\theta})$ \\
\hline
\end{tabular}
\end{table}
One iteration of the proposed Gibbs sampler for sampling from the partially tempered posterior $p(\vec{\theta},\vec{\gamma},\sigma_v^2|\vec{y};T) \propto p(\vec{y}|\vec{\theta},\vec{\gamma},\sigma_v^2)^{1/T}p(\vec{\theta},\vec{\gamma},\sigma_v^2)$ for a given temperature $T$ is given in Table \ref{tab_proposed_gibbs_sampler}. This is a valid Gibbs sampler, which samples each variable at least once within one iteration. The validity of the sampler is established in Section I of the supplementary material by showing that the MH acceptance probability is always 1 for each step. Here, due to our selection of conjugate priors for $\sigma_v^2$ and $\vec{\gamma}$, the partially tempered posterior conditionals $p(\sigma_v^2|\vec{y},\vec{\theta},\vec{\gamma};T) \propto p(\vec{y}|\vec{\theta},\vec{\gamma},\sigma_v^2)^{1/T}p(\sigma_v^2)$ and  $p(\vec{\gamma}|\vec{y},\vec{\theta},\sigma_v^2;T) \propto p(\vec{y}|\vec{\theta},\vec{\gamma},\sigma_v^2)^{1/T}p(\vec{\gamma})$ in Steps 1 and 2 have well-known forms in which the sampling is straightforward. However, the posterior conditional of the multilayer model parameters $p(\vec{\theta}|\vec{y},\vec{\gamma},\sigma_v^2;T) \propto p(\vec{y}|\vec{\theta},\vec{\gamma},\sigma_v^2)^{1/T}p(\vec{\theta})$, given in Step 3, is highly complex and does not have a well-known form, which prevents direct sampling of $\vec{\theta}$. Therefore, we will create a hierarchical sampling scheme and propose a hybrid sampling mechanism combining Slice Sampling and Hamiltonian Monte Carlo approaches, to draw samples from $p(\vec{\theta}|\vec{y},\vec{\gamma},\sigma_v^2;T)$. We present the details of the proposed hybrid sampling method in Section \ref{sec_prop_hybrid_sampler}. We now describe how the Parallel Tempering approach is incorporated with the proposed Gibbs sampler, followed by the derivation of sampling distributions for Steps 1 and 2.\par

Considering a Parallel Tempering scheme with $L$ temperature levels, each MCMC chain samples from a specific partially tempered version of the posterior distribution, i.e., the chain at level $T_{\ell}$ samples from $p(\vec{\theta},\vec{\gamma},\sigma_v^2|\vec{y};T_{\ell}) \propto p(\vec{y}|\vec{\theta},\vec{\gamma},\sigma_v^2)^{1/T_{\ell}}p(\vec{\theta},\vec{\gamma},\sigma_v^2)$ for $\ell = 1,2,\hdots,L$. After one iteration of the Gibbs sampler is completed at all chains, a parameter exchange between the neighbouring levels, say, $T_{\ell}$ and $T_{\ell+1}$, is proposed, where $\ell$ is randomly selected from the uniformly distributed proposal distribution $q_\ell = 1/(L-1)$ for $\ell \in \{1,2,\hdots,L-1\}$. The proposal is accepted with the following acceptance probability
\begin{equation}\label{pt_temp_swap_acc}
    \alpha_{\ell} = \min\bigg\{1,\dfrac{p(\vec{y}|\vec{\theta}^{(\ell,j)},\vec{\gamma}^{(\ell,j)},\sigma_v^{2(\ell,j)})^{1/T_{\ell+1}-1/T_{\ell}}}{p(\vec{y}|\vec{\theta}^{(\ell+1,j)},\vec{\gamma}^{(\ell+1,j)},\sigma_v^{2(\ell+1,j)})^{1/T_{\ell+1}-1/T_{\ell}}}\bigg\},
\end{equation}
where $(\vec{\theta}^{(\ell,j)},\vec{\gamma}^{(\ell,j)},\sigma_v^{2(\ell,j)})$ and $(\vec{\theta}^{(\ell+1,j)},\vec{\gamma}^{(\ell+1,j)},\sigma_v^{2(\ell+1,j)})$ represent the current parameter values at $j^{th}$ MCMC iteration which are to be exchanged between the chains running at level $T_{\ell}$ and $T_{\ell+1}$ respectively (See Section II of the supplementary material for derivation of the acceptance probability). Therefore, one complete MCMC cycle consists of $L$ regular Gibbs sampling stages, followed by a single parameter exchange step. Each cycle $j$ produces a new set of samples for each temperature level, $\{(\vec{\theta}^{(\ell,j)},\vec{\gamma}^{(\ell,j)},\sigma_v^{2(\ell,j)})\}_{\ell=1}^{L}$, but in the end, we are only interested in the samples generated at the first level, $T_1 = 1$, which corresponds to the original posterior distribution. We provide a more detailed description of the sampler in Algorithm \ref{algo_1}. Next, we present the sampling distributions for the first two steps of our sampler, associated with each temperature level. The derivations are provided in Section III of the supplementary material.\par

\subsubsection{Sampling Distribution for Step 1}
The partially tempered posterior conditional distribution for the noise variance $\sigma_v^2$ for a given temperature level $T$ is given by $p(\sigma_v^2|\vec{y},\vec{\theta},\vec{\gamma};T) = \mathcal{IG}(\sigma_v^2;\tilde{\alpha}_v,\tilde{\beta}_v)$ with $\tilde{\alpha}_v = \alpha_v + N/T$ and $\tilde{\beta}_v = \beta_v + \|\vec{y} - \text{diag}(\vec{B}\vec{\gamma})\vec{x}\|^2/T$. Sampling $\sigma_v^2$ is straightforward due to its well-known sampling distribution. Note that as $T \rightarrow \infty$, we have $\tilde{\alpha}_v \rightarrow \alpha_v$ and $\tilde{\beta}_v \rightarrow \beta_v$, which corresponds to the prior distribution $p(\sigma_v^2)$.

\subsubsection{Sampling Distribution for Step 2}
This step requires the partially tempered posterior conditional of the pulse coefficient $\vec{\gamma}$ for a given temperature level $T$, which has the form of a multivariate Gaussian law: $p(\vec{\gamma}|\vec{y},\vec{\theta},\sigma_v^2;T) = \mathcal{N}(\vec{\gamma};\tilde{\vec{\mu}}_{\gamma},\tilde{\vec{\Sigma}}_{\gamma})$
with $\tilde{\vec{\mu}}_{\gamma} = \frac{2}{T\sigma_v^2}\tilde{\vec{\Sigma}}_{\gamma}\Re\{\vec{C}^H\vec{y}\}$ and $\tilde{\vec{\Sigma}}_{\gamma} = \big(\frac{2}{T\sigma_v^2}\Re\{\vec{C}^H\vec{C}\} + \vec{\Sigma}_{\gamma}^{-1}\big)^{-1}$ where $\vec{C} = \text{diag}(\vec{x})\vec{B}$ and $\Re\{\cdot\}$ denotes the real part of its argument. Hence sampling $\vec{\gamma}$ is also straightforward. Similar to Step 1, as $T \rightarrow \infty$, the distribution converges to the prior distribution $p(\vec{\gamma})$ since $\tilde{\vec{\mu}}_{\gamma} \rightarrow \vec{0}$ and $\tilde{\vec{\Sigma}}_{\gamma} \rightarrow \vec{\Sigma}_{\gamma}$.\par

\begin{algorithm}[t]
\DontPrintSemicolon
\SetAlgoLined
\SetInd{3pt}{10pt}
Initialize $\sigma_v^{2(\ell,0)}$, $\vec{\gamma}^{(\ell,0)}$, and $\vec{\theta}^{(\ell,0)}$ for $\ell=1,2,\hdots,L$\;
\For{$j = 1$ \KwTo $J$}{
    \For{$\ell = 1$ \KwTo $L$}{
        Draw $\sigma_v^{2(\ell,j)}$ from $p(\sigma_v^2|\vec{y},\vec{\theta}^{(\ell,j-1)},\vec{\gamma}^{(\ell,j-1)};T_{\ell})$\;
        Draw $\vec{\gamma}^{(\ell,j)}$ from $p(\vec{\gamma}|\vec{y},\vec{\theta}^{(\ell,j-1)},\sigma_v^{2(\ell,j)};T_{\ell})$\;
        Draw $\vec{\theta}^{(\ell,j)}$ from $p(\vec{\theta}|\vec{y},\vec{\gamma}^{(\ell,j)},\sigma_v^{2(\ell,j)};T_{\ell})$\;
    }
    Draw a level $\ell$ uniformly from $\{1,2,\hdots,L-1\}$\;
    Compute acceptance probability $\alpha_{\ell}$ using (\ref{pt_temp_swap_acc})\;
    \If{$U[0,1] < \alpha_\ell$}{
        Swap parameters $\sigma_v^{2(\ell,j)} \rightleftharpoons \sigma_v^{2(\ell+1,j)}$ \;
        Swap parameters $\vec{\gamma}^{(\ell,j)} \rightleftharpoons \vec{\gamma}^{(\ell+1,j)}$\;
        Swap parameters $\vec{\theta}^{(\ell,j)} \rightleftharpoons \vec{\theta}^{(\ell+1,j)}$\;
    }
}
\caption{Proposed Gibbs Sampler with PT\label{algo_1}}
\end{algorithm}

\section{Proposed Hybrid Sampler for Sampling Multilayered Model Parameters} \label{sec_prop_hybrid_sampler}
The multidimensional sampling distribution for the multilayer model parameters $\vec{\theta}$ does not have a well-known form that would enable direct sampling. Therefore, we construct a hierarchical scheme that incorporates a different sampling approach for Step 3 in Table \ref{tab_proposed_gibbs_sampler}. Although PT approach helps resolving the multimodality (or local optimality) issue of the likelihood, the employed sampling scheme still plays an important role on the sampling efficiency. To this end, in this section, we present a specific hybrid sampling mechanism which combines the Slice Sampling (SS) and Hamiltonian Monte Carlo (HMC) approaches. In the following sections, we first describe the principles of SS and HMC, and then present our hybrid sampling scheme.

\subsection{Slice Sampling}\label{subsec_slice}
SS is among the widely used methods for within-Gibbs sampling schemes \cite{RNeal}. It is applicable to both univariate and multivariate cases when the target distribution can be calculated up to a scale. In this work, we employ the univariate setting and sample $\vec{\theta}$ in $3M$ steps, where in each step, we sample an element $\theta_i$ from its full conditional posterior distribution $p(\theta_i|\vec{y},\vec{\theta}_{-i},\vec{\gamma},\sigma_v^2;T)$, associated with a given temperature level $T$. The first step of SS is to randomly draw a density level $\eta_i$ from $U[0,p(\theta_i|\vec{y},\vec{\theta}_{-i},\vec{\gamma},\sigma_v^2;T)]$. Then, a line segment (or a hyper-rectangle for multivariate case) with predefined length, $w_i$, is randomly positioned around $\theta_i$ and sequentially extended in both directions with multiples of $w_i$ until both ends are above $p(\theta_i|\vec{y},\vec{\theta}_{-i},\vec{\gamma},\sigma_v^2;T)$, which is known as the stepping-out procedure. Once the stepping-out procedure is completed, a point  $\tilde{\theta}_i$ is drawn uniformly within the extended line segment. If the selected point does not satisfy $p(\tilde{\theta}_i|\vec{y},\vec{\theta}_{-i},\vec{\gamma},\sigma_v^2;T) \geq \eta_i$, the line segment is shrunk by setting one end to $\tilde{\theta}_i$ such that $\theta_i$ still lies within the resulting line segment and a new point is drawn randomly in the same manner. The shrinkage process, also known as stepping-in procedure, continues until a point satisfies $p(\tilde{\theta}_i|\vec{y},\vec{\theta}_{-i},\vec{\gamma},\sigma_v^2;T) \geq \eta_i$. Once such a point is selected, it is assigned as the next sample value. Throughout this work, we set the length of line segment as the range of corresponding parameter, i.e., $w_i = \theta_{i,\text{max}} - \theta_{i,\text{min}}$.
\begin{figure}[t!]
  \begin{subfigure}[b]{0.242\textwidth}
    \centering
    \includegraphics[width=\textwidth]{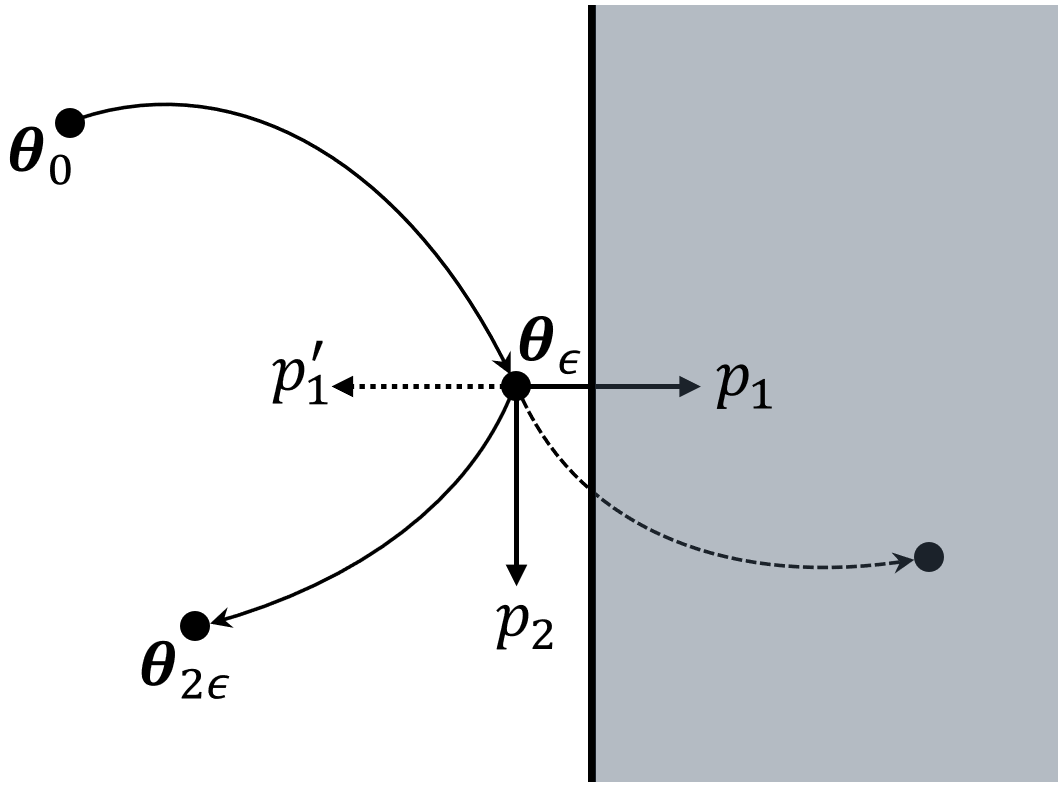}
    \label{fig:hmc_reflection1}
  \end{subfigure}
  \begin{subfigure}[b]{0.238\textwidth}
    \centering
    \includegraphics[width=\textwidth]{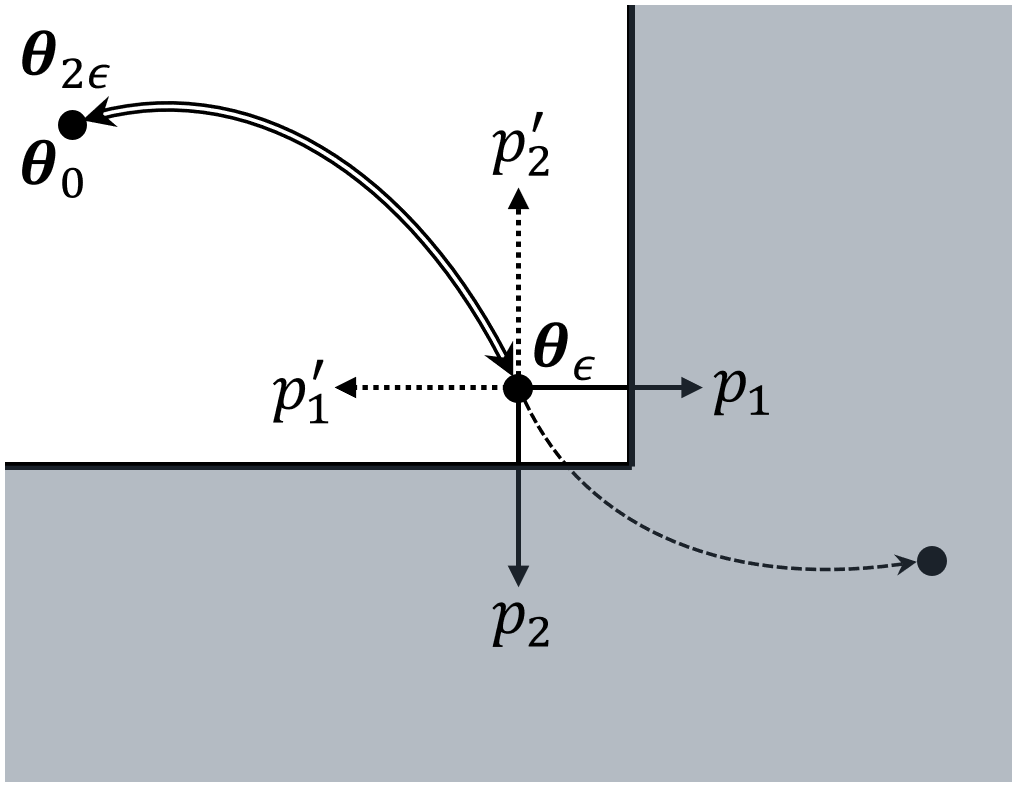}
    \label{fig:hmc_reflection2}
  \end{subfigure}
  \caption{Illustration of reflective HMC for two-dimensional case when (left) only one boundary is violated and (right) both boundaries are violated. Shaded regions represent outside of the boundaries. \label{fig_hmc_with_reflection}}
\end{figure}

\subsection{Hamiltonian Monte Carlo}\label{subsec_hmc}
The core idea of HMC is to utilize the geometry of the target distribution to eliminate the random walk behaviour of the conventional Metropolis-Hastings (MH) method by enabling longer jumps in parameter space with high acceptance rate \cite{RNeal2}. It is based on an analogy with physical systems, in which the target distribution is translated to a potential energy function, where the parameters of interest, $\vec{\theta}$, are regarded as position variables. An augmented state-space is created by introducing momentum variables, denoted by $\vec{p}$, representing the rate of change of the position variables. Defining the tempered potential energy function as $U(\vec{\theta};T) =  -\log p(\vec{y}|\vec{\theta},\vec{\gamma},\sigma_v^2;T)$ and the kinetic energy function as $K(\vec{p}) = \frac{1}{2}\vec{p}^T\vec{M}\vec{p}$, where $\vec{M}$ is a weighting matrix that adjusts the momentum distribution for more efficient sampling, total energy of the system at a given state $(\vec{\theta},\vec{p})$ at temperature $T$ is given by the Hamiltonian $H(\vec{\theta},\vec{p};T) = U(\vec{\theta};T) + K(\vec{p})$. \par

HMC is used to sample $(\vec{\theta},\vec{p})$ pairs jointly from the canonical distribution $P(\vec{\theta},\vec{p};T) \propto \exp\big(-H(\vec{\theta},\vec{p};T)\big)$ at a given temperature level $T$. The sampling is achieved by first sampling a new momentum state from $\mathcal{N}(\vec{p};\vec{0},\vec{M}^{-1})$, and then simulating the Hamiltonian dynamics, given by
\begin{equation}\label{hamiltonian_dynamics}
    \dfrac{d\vec{\theta}}{d t} = \nabla_{p}H(\vec{\theta},\vec{p};T), \qquad \dfrac{d\vec{p}}{d t} = -\nabla_{\theta}H(\vec{\theta},\vec{p};T),
\end{equation}
to produce a new position state. However, exact simulation requires integration of (\ref{hamiltonian_dynamics}), which is not feasible in practice. Hence, it is approximated by the leapfrog algorithm, which is a numerical integration scheme consisting of alternating discretized updates to $\vec{\theta}$ and $\vec{p}$: $i)$ $\vec{p}_{\epsilon/2} = \vec{p}_{0} - \frac{\epsilon}{2}\nabla_{\theta}U(\vec{\theta}_{0};T)$, $ii)$ $\vec{\theta}_{\epsilon} = \vec{\theta}_{0} + \epsilon \vec{M}\vec{p}_{\epsilon/2}$, and $iii)$ $\vec{p}_{\epsilon} = \vec{p}_{\epsilon/2} - \frac{\epsilon}{2}\nabla_{\theta}U(\vec{\theta}_{\epsilon};T)$. \par 

One iteration of the leapfrog algorithm simulates the dynamics for a time interval $\epsilon$, which is the predefined step size of the algorithm. In order to simulate for a duration of $\tau$, the process is repeated for $\Delta = \tau/\epsilon$ times. Although the leapfrog algorithm provides quite accurate approximation of the continuous time integration, some residual error will remain due to discretization, which might alter the value of Hamiltonian. In order to maintain detailed balance, the proposed state is accepted with MH acceptance criterion. \par
\begin{figure}[t!]
    \centering
    \includegraphics[width=1\linewidth]{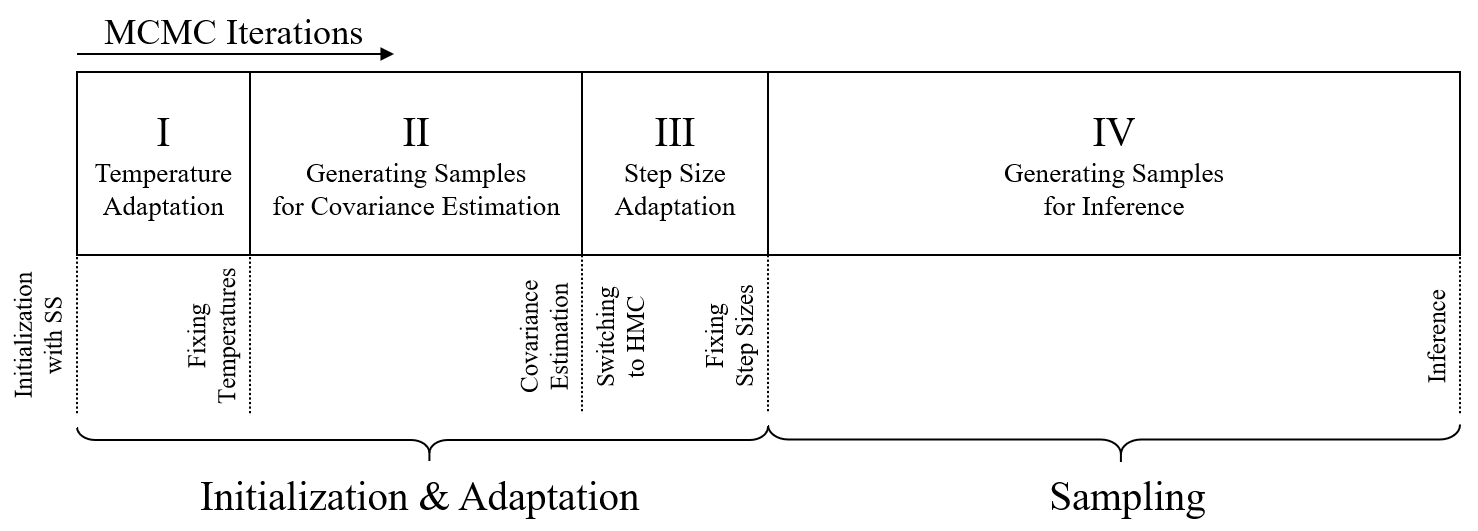}
  \caption{Proposed hybrid sampling mechanism with self-adaptation.\label{fig_hybrid_sampling_scheme}}
\end{figure}
HMC is conventionally used for sampling from smooth and unbounded distributions. For bounded parameter spaces, as we have with $\Lambda_{\theta}$, a modified reflective HMC can be used, where the trajectory on the parameter space is bounced back when it is blocked by a boundary. Specifically, if $\theta_i \notin [\theta_{i,\text{min}},\theta_{i,\text{max}}]$ after completing one step of the leapfrog algorithm, we undo the previous step, negate the $i^{th}$ momentum variable, i.e., $p_{i}^{\prime} = -p_i$, and then complete the remaining steps using the updated momentum vector. If multiple boundaries are violated simultaneously, all of the corresponding momentum variables are negated. In Fig. \ref{fig_hmc_with_reflection}, we demonstrate the employed reflection method for a two-dimensional case. This method of reflection leaves the Hamiltonian invariant, since negation does not change the value of kinetic energy function, i.e., $K(\vec{p}^{\prime}) = K(\vec{p})$. Moreover, the same MH acceptance criterion remains valid, preserving the detailed balance. \par 
\begin{figure*}[t!]
\centering
  \begin{subfigure}[b]{0.49\textwidth}
  \centering
    \includegraphics[width=0.85\textwidth]{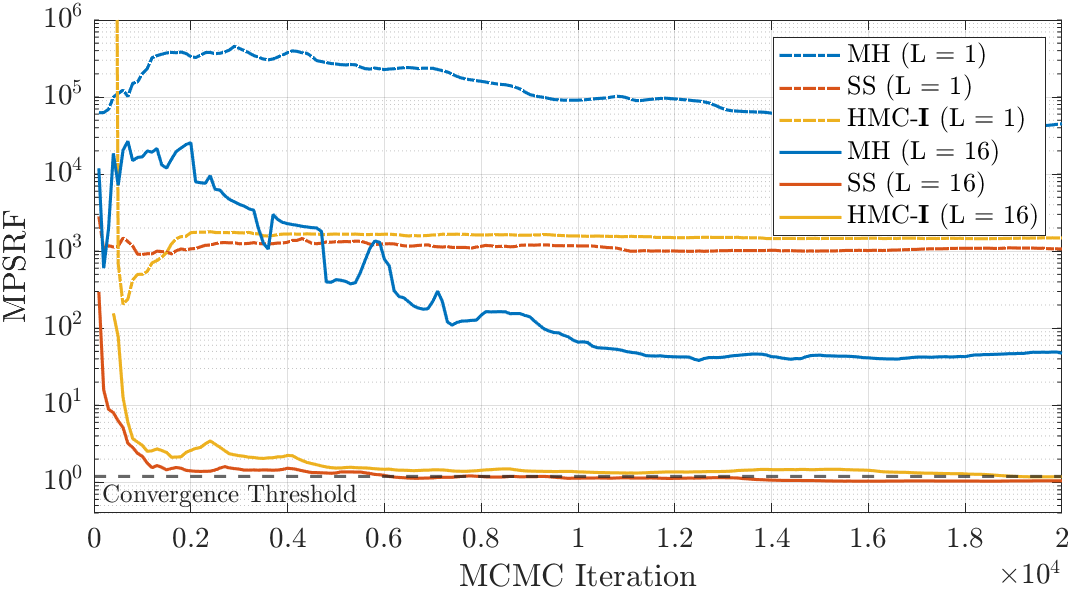}
  \end{subfigure}
  \begin{subfigure}[b]{0.49\textwidth}
   \centering
    \includegraphics[width=0.85\textwidth]{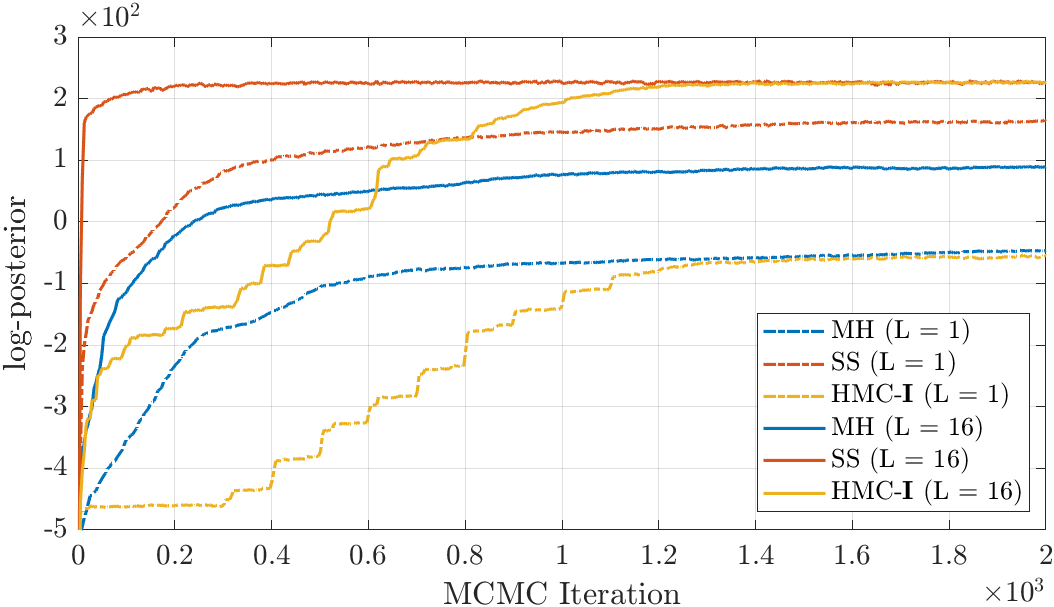}
  \end{subfigure}
  \caption{Evolution of MPSRF and log-posterior for different samplers with $L=1$ (No Tempering) and $L=16$ (Parallel Tempering). \label{fig_convergence_comparison}}
\end{figure*}
Note that the leapfrog algorithm still requires the analytic expression for the gradient of the potential energy function $U(\vec{\theta};T) = \|\vec{y} - \text{diag}(\vec{B}\vec{\gamma})\vec{x}\|^2/T\sigma_v^2$. Following the derivation provided in Section IV of the supplementary material, it is given by
\begin{equation}\label{gradient_potential}
    \nabla_{\theta}U(\vec{\theta})
    = -\dfrac{2}{T\sigma_v^2}\Re\Big\{\big(\vec{y} - \vec{D}\vec{x}\big)^{H}\vec{D}\nabla_{\theta}\vec{x}\Big\},
\end{equation}
where $\vec{D} = \text{diag}(\vec{B}\vec{\gamma})$. The gradient of $\vec{x}$ is defined as $\nabla_{\theta}\vec{x} = [\nabla_{\theta}X_0(\omega_0),\nabla_{\theta}X_0(\omega_1),\hdots,\nabla_{\theta}X_0(\omega_{N-1})]^T$, where the individual terms $\nabla_{\theta}X_0(\omega_i)$ have the form of $\nabla_{\theta}X_0(\omega_i) = [\partial X_0(\omega_i)/\partial\theta_1,\hdots,\partial X_0(\omega_i)/\partial\theta_{3M}]^T$
for $i =0,1,\hdots,N-1$. Exact expression for each element of $\nabla_{\theta}X_0(\omega_i)$ is also provided in Section IV of the supplementary material.

\subsection{Proposed Hybrid Sampler with Self-Adaptation}
The parameters $\epsilon$, $\Delta$ and $\vec{M}$ affect the overall performance of HMC significantly. In general, higher $\epsilon$ causes high residual error leading to low acceptance rate. On the other hand, selecting a too small $\epsilon$ will require large number of steps $\Delta$ to achieve long jumps, which increases the computational load. Hence, both parameters need to be tuned for the best trade-off. Similarly, appropriate selection of $\vec{M}$ is crucial for sampling efficiency. Note that the residual error is actually the sum of all errors in each dimensions. Therefore, the simple choice of $\vec{M} = \vec{I}$, which assigns equal weights for all dimensions, causes step size $\epsilon$ to be determined according to the dimension with the smallest variance. This is because a smaller variance at a given direction generally corresponds to a higher gradient in that direction, which increases sensitivity to the value of the momentum. The performance can be significantly improved by adjusting the momentum variables accordingly to maintain a similar level of error in each dimension. This can be achieved by selecting $\vec{M}$ as a diagonal matrix consisting of the inverse of the variances in each dimension. A better strategy would be to set $\vec{M}$ as the inverse of the full covariance matrix $\vec{\Sigma}$, which would not only incorporate the variance information, but also capture the linear correlations between the parameters. However, for complex distributions, analytical calculation of the covariance matrix is not possible, and hence, an estimate is required. In addition to these issues, another essential but non-trivial problem is the selection of the temperature ladder for PT scheme as it has a substantial effect on the overall exploration power of the sampler. Since no unique set of temperature levels exist that works well for all distributions, the temperatures should be adjusted for improved sampling performance. \par 

To address the issues described above, we designed an adaptive sampling mechanism that consists of an initialization/adaptation stage as part of the burn-in process for learning the temperature levels as well as the covariance matrices and the step sizes (for fixed number of steps $\Delta$) associated with each temperature level from the measurement. As illustrated in Fig. \ref{fig_hybrid_sampling_scheme}, we initialize the sampling process using SS and iteratively learn the temperature levels in Stage I through the mechanism described in Section \ref{subsec_adaptive_temp}. Once a certain convergence criterion is satisfied, we fix the temperatures and start generating samples for the covariance estimation in Stage II. After having the covariance estimates for each temperature level, in Stage III, we switch to HMC approach, set $\vec{M} = \vec{\hat{\Sigma}}_{SS}^{-1}$ and learn the step sizes associated with each temperature level in a sequential manner as described in Section \ref{subsec_adaptive_step_size}. After convergence, we fix the step sizes and start the actual sampling process for inference. \par 

The proposed sampling mechanism combines the SS and HMC approaches, yielding a hybrid model. Our main motivations for initializing the process with SS and then switching the HMC are as follows: Firstly, we only need to set the widths of the hyper-rectangles for SS, which, as our experiments indicated, does not have a crucial effect on the sampling performance and can be fixed from the initialization. This creates a controlled sampling period for more accurate temperature adjustment. Secondly, SS achieves the fastest convergence rate compared to conventional MH and HMC that uses an identity weight matrix, as we will illustrate in Section \ref{sec_simulations}. Finally, HMC achieves outstanding sampling efficiency after convergence if the weighting matrix is well-adjusted to capture the correlations between different parameters. Therefore, the idea is to combine the convergence speed of SS with the sampling efficiency of HMC to create a more powerful sampling method. In the following sections, we describe the adaptive models employed in Stage I and III for temperature level and step size adjustments. \par
\begin{table*}[t!]
\centering
\caption{Autocorrelation time (ACT) of the samplers for each model parameter. Lowest value in each column is represented in bold.}
\label{tab_ACT_comparison}
\resizebox{\textwidth}{!}{%
\begin{tabular}{@{}cccccccccccccccc@{}}
\toprule
 & \multicolumn{15}{c}{Model Parameters} \\ \cmidrule(l){2-16} 
Samplers & $\varepsilon_1$ & $\varepsilon_2$ & $\varepsilon_3$ & $\varepsilon_4$ & $\varepsilon_5$ & $\sigma_1$ & $\sigma_2$ & $\sigma_3$ & $\sigma_4$ & $\sigma_5$ & $d_0$ & $d_1$ & $d_2$ & $d_3$ & $d_4$ \\ \midrule
MH & 1272 & 653 & 1092 & 1416 & 2702 & 2521 & 1628 & 1433 & 3250 & 284 & 310 & 1161 & 597 & 1363 & 2406 \\
SS & 1026 & 516 & 965 & 757 & 106 & 148 & 628 & 135 & 30 & 60 & 164 & 1018 & 484 & 957 & 691 \\
HMC-\textbf{\textbf{I}} & 510 & 409 & 750 & 1279 & 728 & 704 & 488 & 639 & 1007 & 1262 & 238 & 502 & 403 & 750 & 1295 \\
HMC-\textbf{$\vec{\hat{\Sigma}}_{SS}$} & \textbf{56} & \textbf{63} & \textbf{35} & \textbf{51} & \textbf{69} & \textbf{34} & \textbf{28} & \textbf{64} & \textbf{25} & \textbf{28} & \textbf{87} & \textbf{50} & \textbf{62} & \textbf{34} & \textbf{55} \\ \bottomrule
\end{tabular}%
}
\vspace{-3mm}
\end{table*}

\begin{figure*}[t!]
\centering
  \begin{subfigure}[b]{0.32\textwidth}
  \centering
    \includegraphics[width=\textwidth]{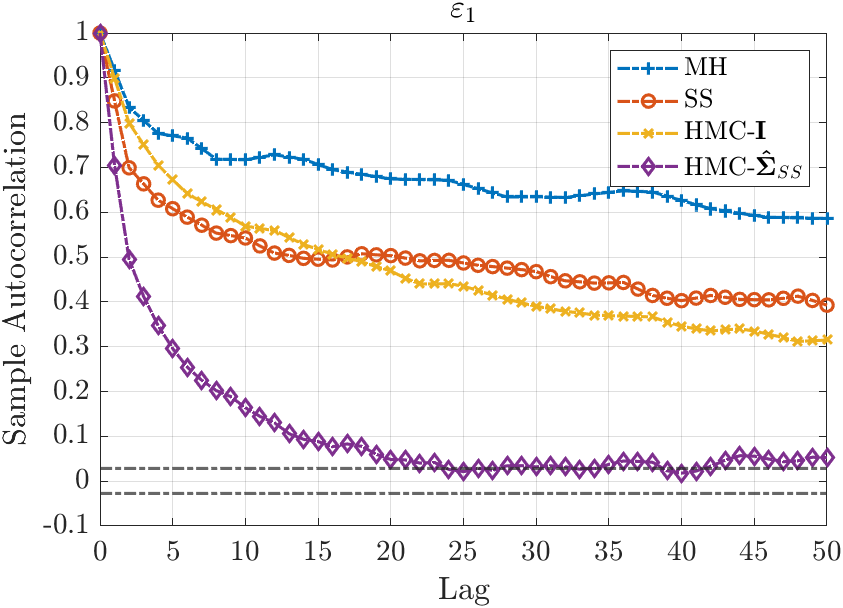}
  \end{subfigure}
  \begin{subfigure}[b]{0.32\textwidth}
   \centering
    \includegraphics[width=1\textwidth]{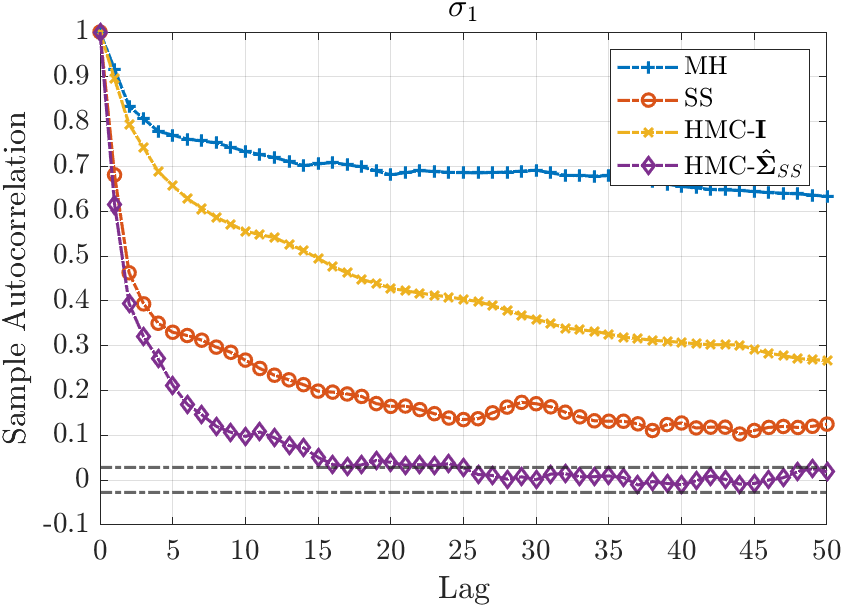}
  \end{subfigure}
  \begin{subfigure}[b]{0.32\textwidth}
   \centering
    \includegraphics[width=1\textwidth]{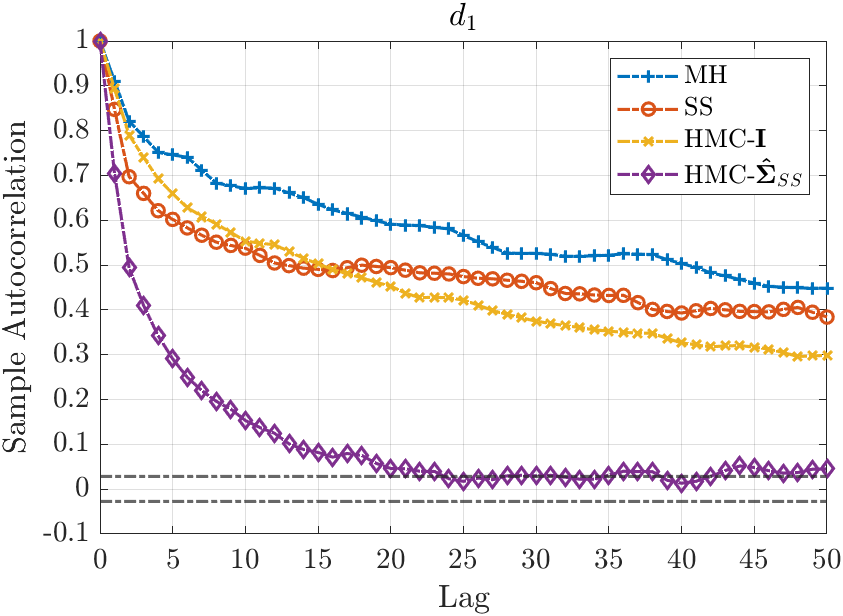}
  \end{subfigure}
  \caption{Autocorrelation functions (ACF) of the samplers for the first layer parameters.\label{fig_ACF_comparison}}
\end{figure*}
\subsubsection{Adaptive Temperature Selection}\label{subsec_adaptive_temp}
For PT, selection of the temperature ladder $T_1<\hdots<T_{L}$ has a substantial effect on the overall sampling performance. The general practice is to set $T_1 = 1$ to sample from the original target distribution and $T_L$ sufficiently high to explore all the modes. There exist different point of views to optimize the structure of the temperature ladder. In this work, we assume that the total number of temperatures is fixed and determined by the available computational budget. It has been shown in the literature that a reasonable approach is to set the temperature spacing such that the swap ratios approximately equal for adjacent levels \cite{VousdenW}. Following this approach, we provide an adaptive temperature selection scheme that iteratively adjusts the temperature levels until a certain convergence criterion is met. Consider an intermediate temperature ladder configuration $\{T_{\ell}^{(j)}\}_{\ell=1}^{L}$ at $j^{th}$ MCMC iteration. The effect of any changes on $\{T_{\ell}^{(j)}\}_{\ell=1}^{L}$ can only be observed in the proceeding iterations. Therefore, we update the temperatures after every $J_T$ iterations based on the empirical swap ratio $s_{\ell}^{(j)}$, which is calculated by the ratio of the total accepted swaps to the total proposed swaps between chains $\ell$ and $\ell+1$ during the iterations $(j-J_T+1)$ and $j$. In order to maintain the order, i.e., $T_1 < \hdots < T_{L}$, and level out the scaling differences, we perform the updates on the logarithm of their difference as 
\begin{equation}
    T_{\Delta_{\ell}}^{(j+1)} = T_{\Delta_{\ell}}^{(j)} - e_{\ell}^{(j)}K_T\mathbbm{1}_{J_T}(j)
\end{equation}
where $T_{\Delta_{\ell}}^{(j)} = \log\big(T_{\ell+1}^{(j)} - T_{\ell}^{(j)}\big)$, $e_{\ell}^{(j)} = s_{\ell+1}^{(j)} - s_{\ell}^{(j)}$, $K_T$ is the controller gain, and $\mathbbm{1}_{J_T}(j)$ refers to the indicator function defined as $\mathbbm{1}_{J_T}(j) = 1$ if $j \bmod J_T = 0$ and $\mathbbm{1}_{J_T}(j) = 0$ otherwise. The initial configuration is generally selected as $L$ geometrically spaced levels between $T_1$ and $T_L$. Here, we note that any adjustment on the temperature levels during the sampling process violates the detailed balance. Therefore, we finalize the temperature update when the variation within the last $N_T$ updates is less than 10\% simultaneously for all levels:
\begin{equation}\label{convergence_criterion_for_adaptation}
    \dfrac{\sqrt{\frac{1}{N_T - 1}\sum_{i=0}^{N_T-1} \big(T_{\ell}^{(j-iJ_T)} - \bar{T}_{\ell}\big)^2}}{\bar{T}_{\ell}} \leq 0.1,
\end{equation}
where $\bar{T}_{\ell} = \frac{1}{N_T}\sum_{i=0}^{N_T-1}T_{\ell}^{(j-iJ_T)}$. We then fix the temperatures and initiate Stage II for covariance estimation.

\subsubsection{Adaptive Step Size Selection}\label{subsec_adaptive_step_size}
In this section, we provide an adaptive model to be used in Stage III, by which we periodically update the step sizes to achieve a predetermined acceptance ratio $\xi$ based on the current empirical acceptance ratios. Similar to temperature adjustments, we update the step sizes after every $J_{\epsilon}$ iterations based on the empirical acceptance ratio $\hat{\xi}_{\ell}^{(j)}$ measured by the ratio of the total accepted proposals between iterations $(j-J_{\epsilon}+1)$ and $j$ to the duration $J_{\epsilon}$. We employ a proportional controller approach and use the difference between the target and empirically measured acceptance ratios, i.e., $e_{\ell}^{(j)} = \xi - \hat{\xi}_{\ell}^{(j)}$, as the model feedback. Hence, the adaptive model is described by
\begin{equation}\label{step_size_adaptive_model}
    \epsilon_{\ell}^{(j+1)} = \exp\big(\log(\epsilon_{\ell}^{(j)}) - e_{\ell}^{(j)}K_{\epsilon}\mathbbm{1}_{J_{\epsilon}}(j)\big),
\end{equation}
where we perform the updates on the logarithm of parameters to level out scale differences and use the same constant gain $K_{\epsilon}$ for all temperature levels. We employ the same convergence criterion defined in (\ref{convergence_criterion_for_adaptation}) and fix the step sizes before initiating Stage IV. As a result, no adaptation is performed and all parameters are fixed during the actual sampling period, which maintains the Markovianity and detailed balance.

\section{Simulations}\label{sec_simulations}
In the first part of this section, we justify our reasoning behind the construction of proposed hybrid sampling mechanism and demonstrate the obtained superior sampling efficiency. Then, in the second part, we investigate the recovery of multilayer model parameters from synthetic measurements simulating human tissues. Throughout this section, we will use MH to denote the Metropolis-Hastings sampling scheme. Since the parameter space is bounded, we use independent Beta distributions for each dimensions as the proposal distribution. We locate the mode at the current sample value and employ the same adaptation model given in (\ref{step_size_adaptive_model}) for the concentration of proposal distributions to achieve a predetermined acceptance rate. Same as before, SS and HMC will represent the Slice Sampling and Hamiltonian Monte Carlo approaches described in Section \ref{subsec_slice} and \ref{subsec_hmc}. More specifically, we will use HMC-\textbf{I} and HMC-$\vec{\hat{\Sigma}}_{SS}$ to denote $\vec{M} = \vec{I}$ and $\vec{M} = \vec{\hat{\Sigma}}_{SS}^{-1}$ cases respectively. In other words, HMC-$\vec{\hat{\Sigma}}_{SS}$ corresponds to the Stage III and IV of the proposed hybrid sampler.\par 

For the experiments, the parameters of prior distributions were selected as $\sigma_{\gamma}^2 = 10$, $\alpha_v = 10^{-3}$, and $\beta_v = 10^{-3}$, which constitute non-informative priors. The subspace matrix $\vec{A}$ for the transmitted waveform was constructed by the first 8 length-$23$ DPS sequences, which span the frequency range of $0$ to $16$ GHz. The lower and upper bounds of the multilayer model parameters were specified as $\varepsilon_{\text{min}} = 2$, $\varepsilon_{\text{max}} = 100$, $\sigma_{\text{min}} = 5 \times 10^{-3}$, $\sigma_{\text{max}} = 3$, $d_{\text{min}} = 10^{-3}$ and $d_{\text{max}} = 3\times10^{-2}$. The associated prior distributions were selected such that the mode $\lambda_i$ is located at the normalized typical value of the corresponding model parameter with concentration $\kappa_i = 100$, except for the last layer parameters, which were assigned flat priors with $\kappa_i = 0$. For parallel tempering, a total of $L = 16$ different temperature levels, initialized at geometrically spaced points in between $T_1 = 1$ and $T_{16} = 10^5$, were employed. We performed temperature updates after every $J_T = 200$ iterations with $K_T = 10$ and $N_T = 10$. For HMC, the step sizes were initialized at $10^{-2}$ with $\xi = 0.85$, $J_{\epsilon} = 100$, and $K_{\varepsilon} = 0.5$. 
\begin{figure*}[t!]
\centering
  \begin{subfigure}[b]{0.49\linewidth}
  \centering
    \includegraphics[width=1\linewidth]{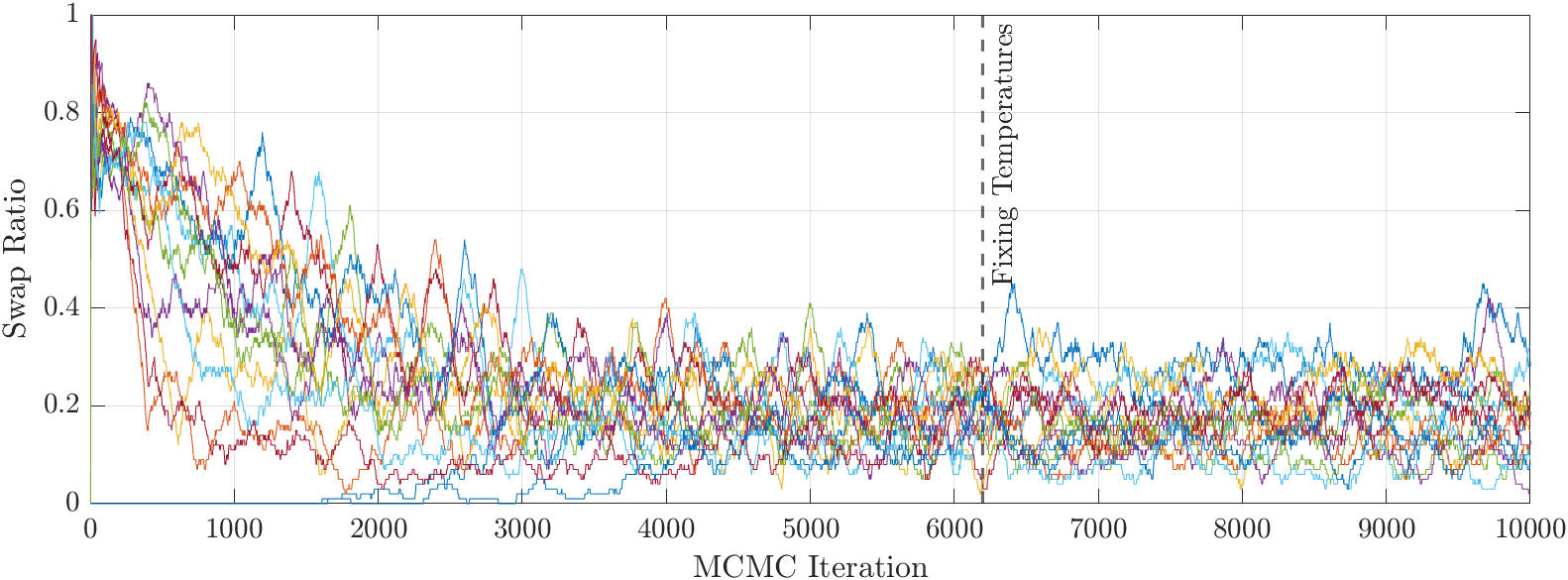}
  \end{subfigure}
  \begin{subfigure}[b]{0.49\linewidth}
  \centering
    \includegraphics[width=1\linewidth]{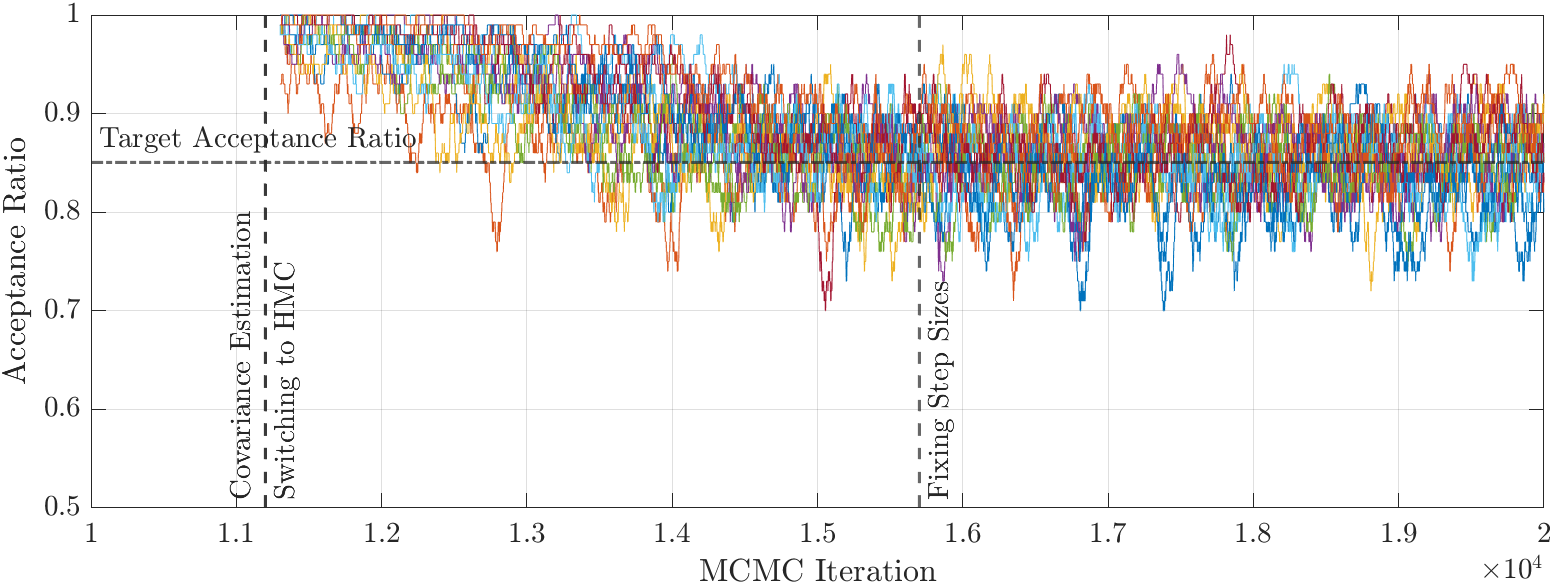}
  \end{subfigure}
    \begin{subfigure}[b]{0.49\linewidth}
  \centering
    \includegraphics[width=1\linewidth]{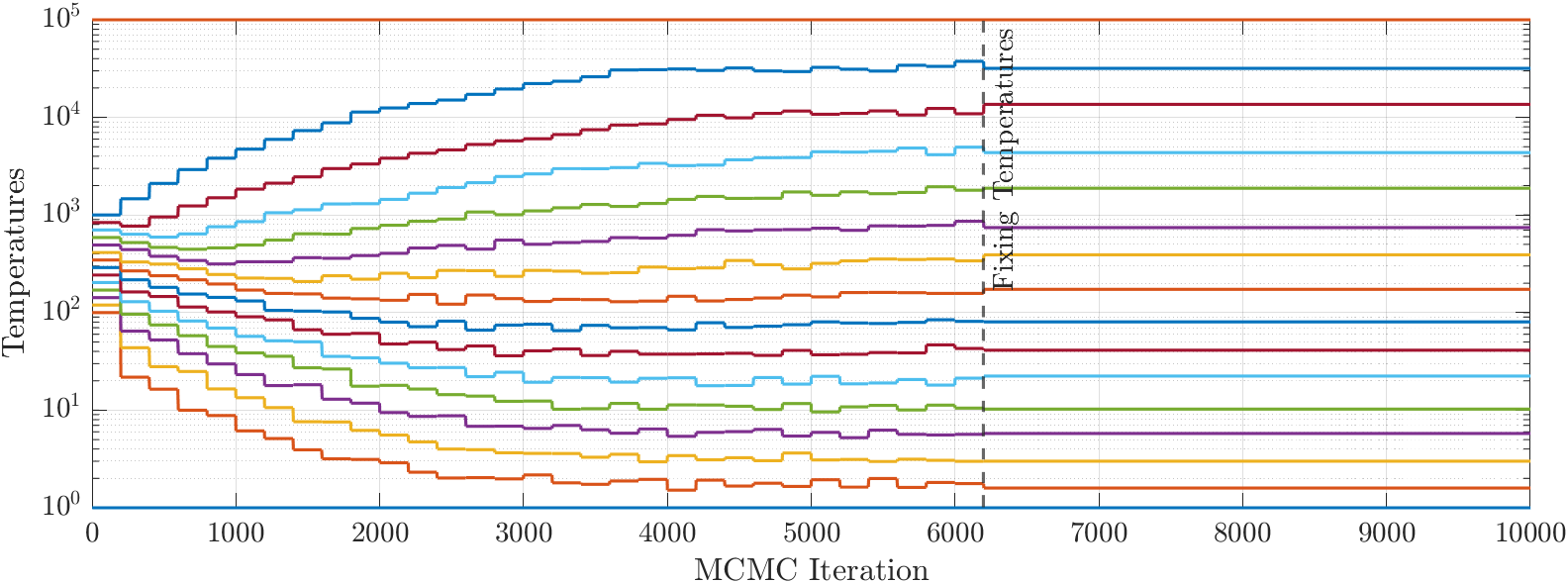}
  \end{subfigure}
  \begin{subfigure}[b]{0.49\linewidth}
  \centering
    \includegraphics[width=1\linewidth]{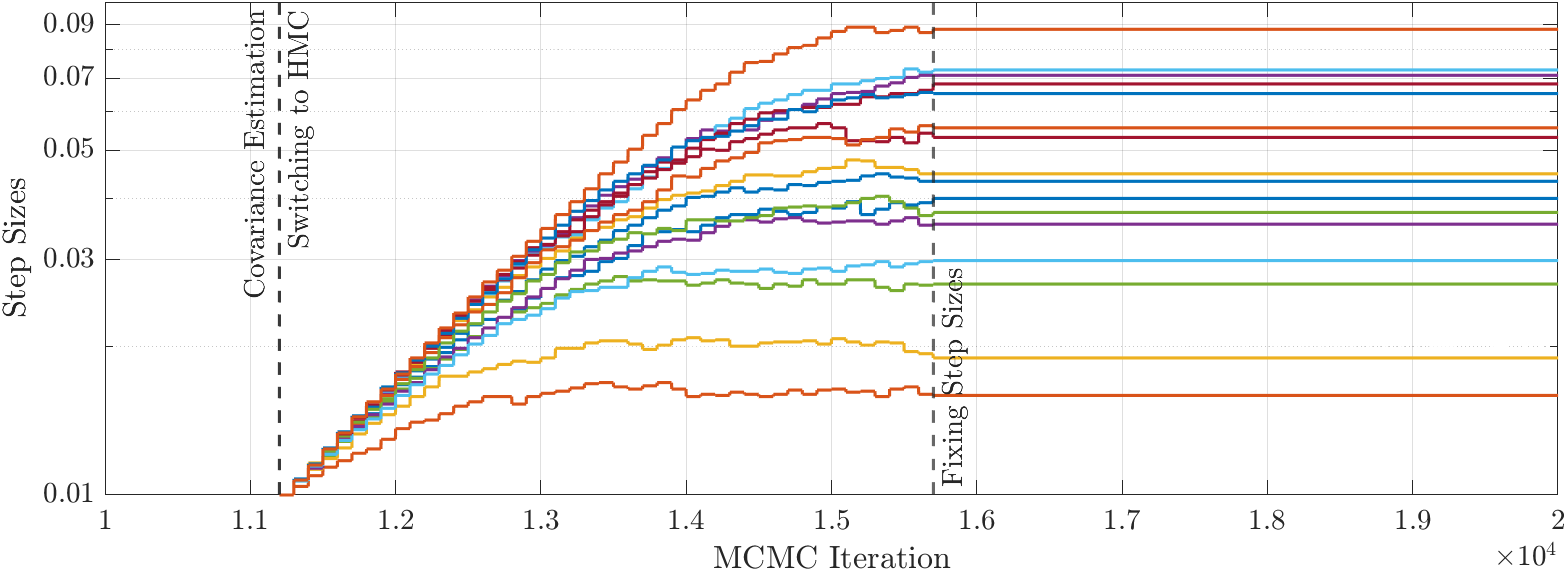}
  \end{subfigure}
  \caption{Evolution of swap ratios (top left) and temperature levels (bottom left) using the adaptive temperature adjustment model with $L = 16$ levels. The lowest and highest temperature levels are fixed at $T_1 = 1$ and $T_{16} = 10^5$. Evolution of acceptance ratios (top right) and step sizes (bottom right) using the adaptive step size adjustment model with target acceptance ratio $\xi = 0.85$. \label{fig_adaptive_adjustment_schemes}}
\end{figure*}
\begin{figure}[t!]
\centering
  \begin{subfigure}[b]{1\linewidth}
    \includegraphics[width=\linewidth]{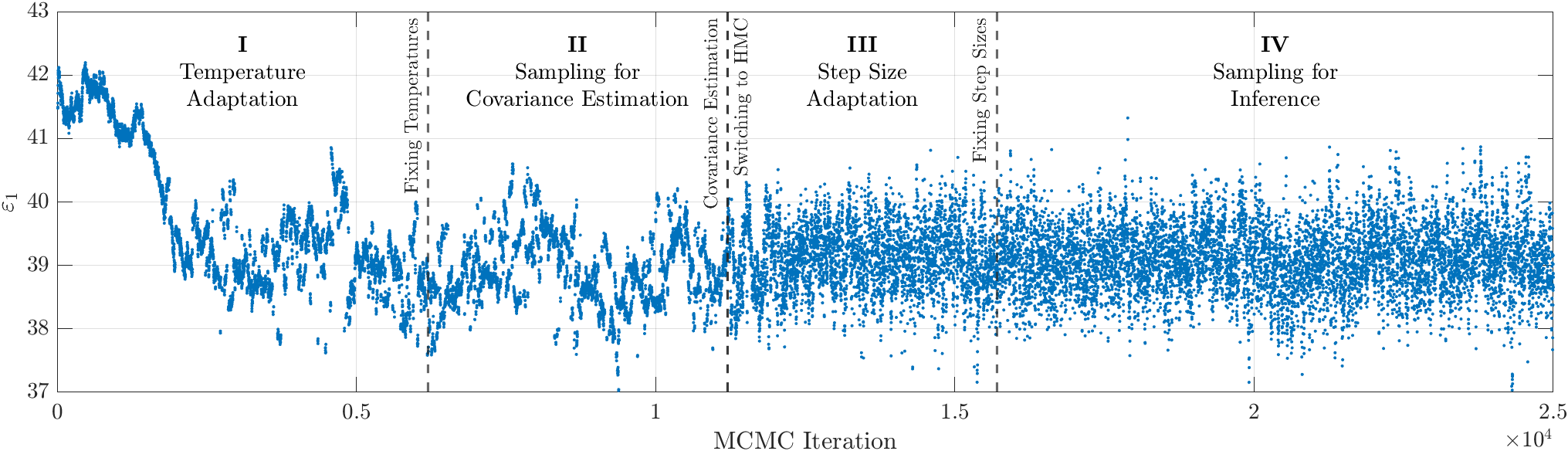}
  \end{subfigure}
  \begin{subfigure}[b]{1\linewidth}
    \includegraphics[width=1\linewidth]{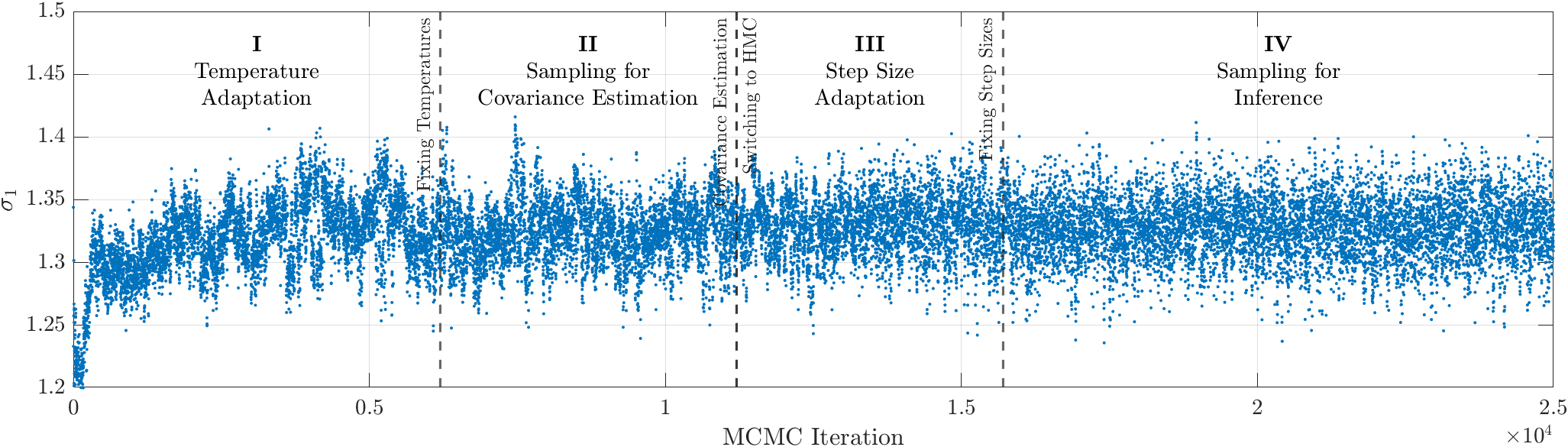}
  \end{subfigure}
  \begin{subfigure}[b]{1\linewidth}
    \includegraphics[width=1\linewidth]{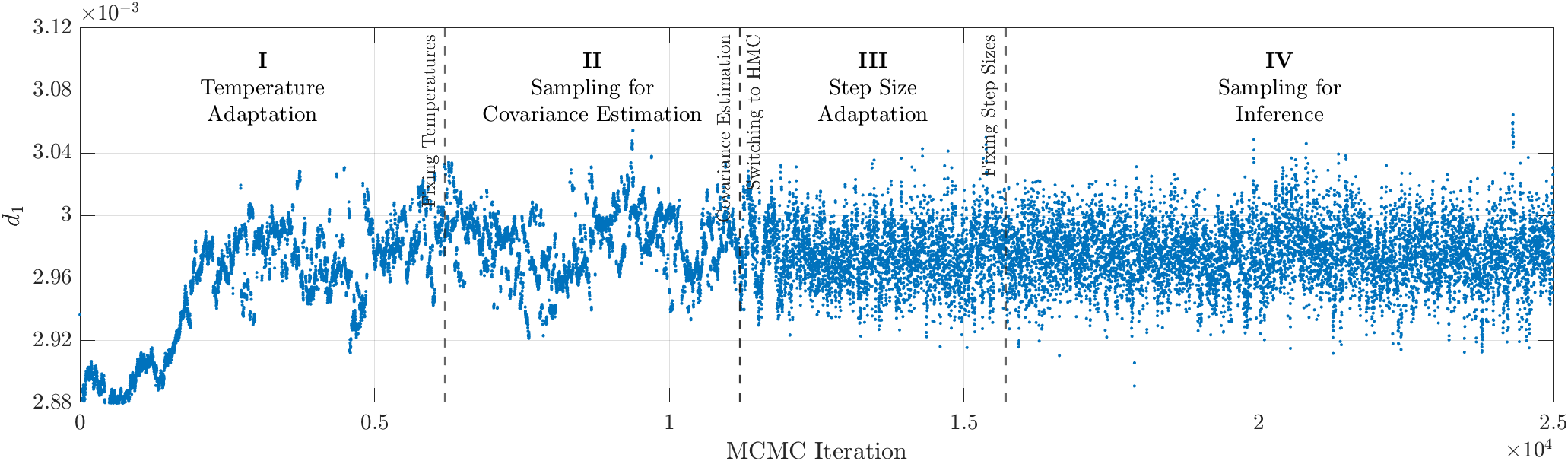}
  \end{subfigure}
  \caption{Trace plots for the parameters $\varepsilon_1$ (top), $\sigma_1$ (middle), and $d_1$ (bottom) at all stages of the proposed sampling procedure. \label{fig_sampling_illustration}}
\end{figure}

\subsection{Convergence Rate Analysis}
One of the main reasons for using SS within the first two stages of our sampling mechanism is its faster convergence rate compared to MH and HMC-\textbf{I}. In this section, we establish this by comparing the empirically measured convergence rates. Since no covariance estimate is available initially, we do not consider HMC-$\vec{\hat{\Sigma}}_{SS}$ for comparison. In order to empirically compare the convergence rates, we first consider the iterated graphical monitoring approach proposed by Brook and Gelman in \cite{SBrooks}. The convergence is measured based on the Multivariate Potential Scale Reduction Factor (MPSRF) as defined in \cite{SBrooks}, which is calculated on multiple simulations running simultaneously and independently. The convergence is declared when MPSRF is close to 1, a typical threshold being 1.2 as suggested in \cite{SBrooks}. \par 

To produce the MPSRF curves, we run 8 different simulations on the same measurement and calculate the MPSRF value after every 100 iterations by using only the second half of the generated samples, where the first half is discarded as part of the burn-in process. Note that we employ a PT scheme and have multiple chains associated with each of these 8 simulations. Since we are only interested in the samples corresponding to $T_1 = 1$, we calculate the MPSRF curves on the first chains. In order to demonstrate the effect of PT, we also considered the scenario in which we do not employ any tempering approach and run a single chain at $T = 1$ for each simulations. The resulting curves are illustrated in Fig. \ref{fig_convergence_comparison}. \par

Our first observation is that PT improves the convergence rates significantly for all samplers. Without PT, the samplers quickly get stuck on a local optimal region depending on their initialization and the MPSRF fails to decrease within the simulation duration. On the other hand, the MPSRF curves for the samplers with PT quickly converge to 1 for both SS and HMC-\textbf{I}. Even though it does not converge as quickly for MH, a significant improvement still exists. This deficiency mainly results from the random walk behaviour of MH, which is inevitable in most complex multivariate distributions without accurate estimation of the curvature information. Comparing SS and HMC-\textbf{I}, although they both get close to 1 very rapidly, it takes, respectively, around 6000 and 19000 iterations for MPSRF to fall below the convergence threshold of 1.2 for SS and HMC-\textbf{I}. This result provides an empirical evidence for the fast convergence rate of SS.\par

As an additional convergence analysis, we also compared the evolution of the value of posterior distribution as simulations progress. We present the mean average of the logarithm of unnormalized posterior value over 8 independent simulations in Fig. \ref{fig_convergence_comparison}. Same as before, the performance improvement obtained via PT scheme is clearly visible for all samplers. Although both SS and HMC-\textbf{I} reach the stationary distribution within the first $2\times10^3$ iterations, SS considerably outperforms HMC-\textbf{I} in terms of the number of iterations needed for convergence, providing another empirical support for selecting SS as the sampling method employed within the first two stages of the proposed sampling mechanism.
\begin{figure*}[t!]
\centering
  \begin{subfigure}[b]{0.32\textwidth}
  \centering
    \includegraphics[width=1\textwidth]{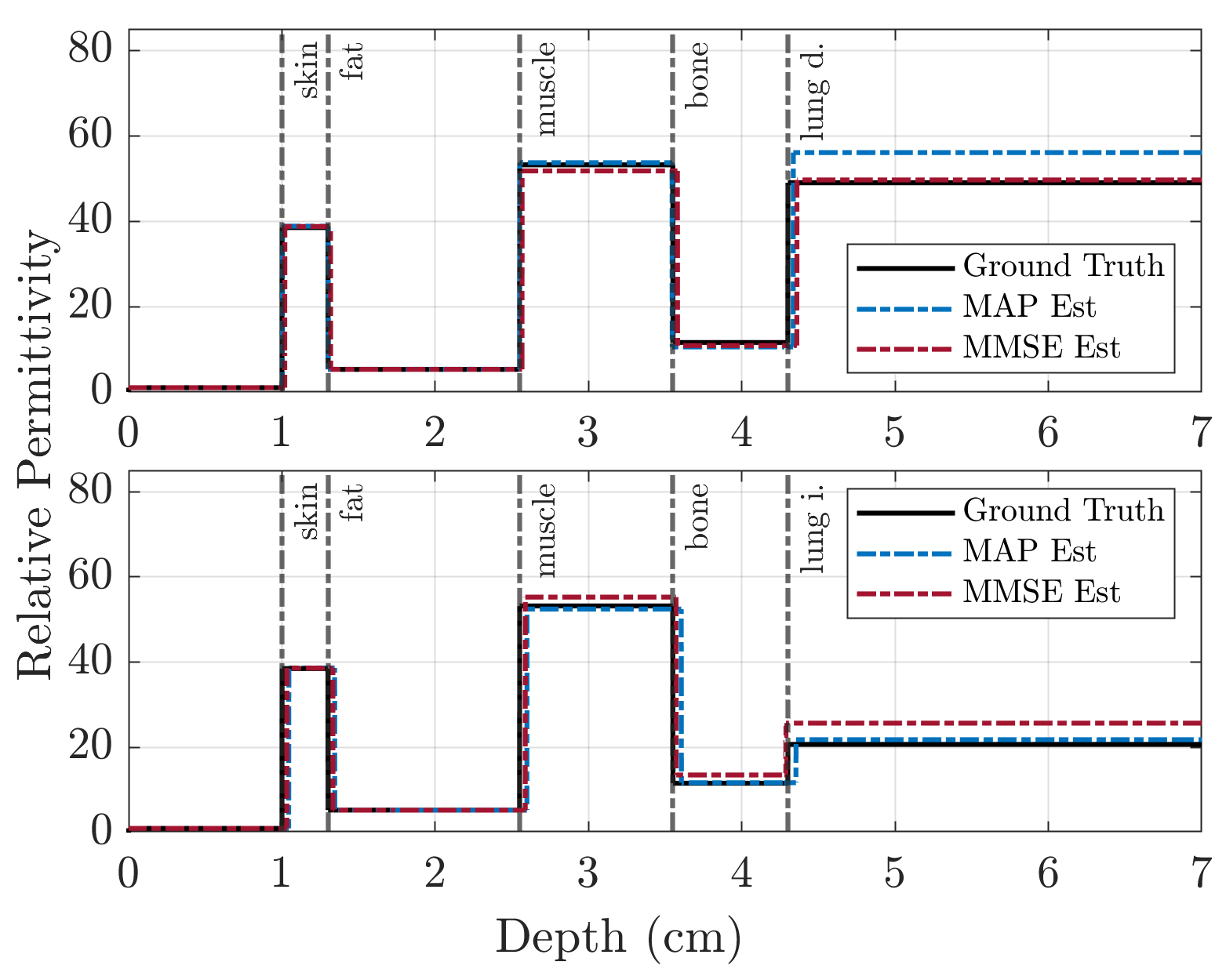}
  \end{subfigure}
  \begin{subfigure}[b]{0.32\textwidth}
  \centering
    \includegraphics[width=1\textwidth]{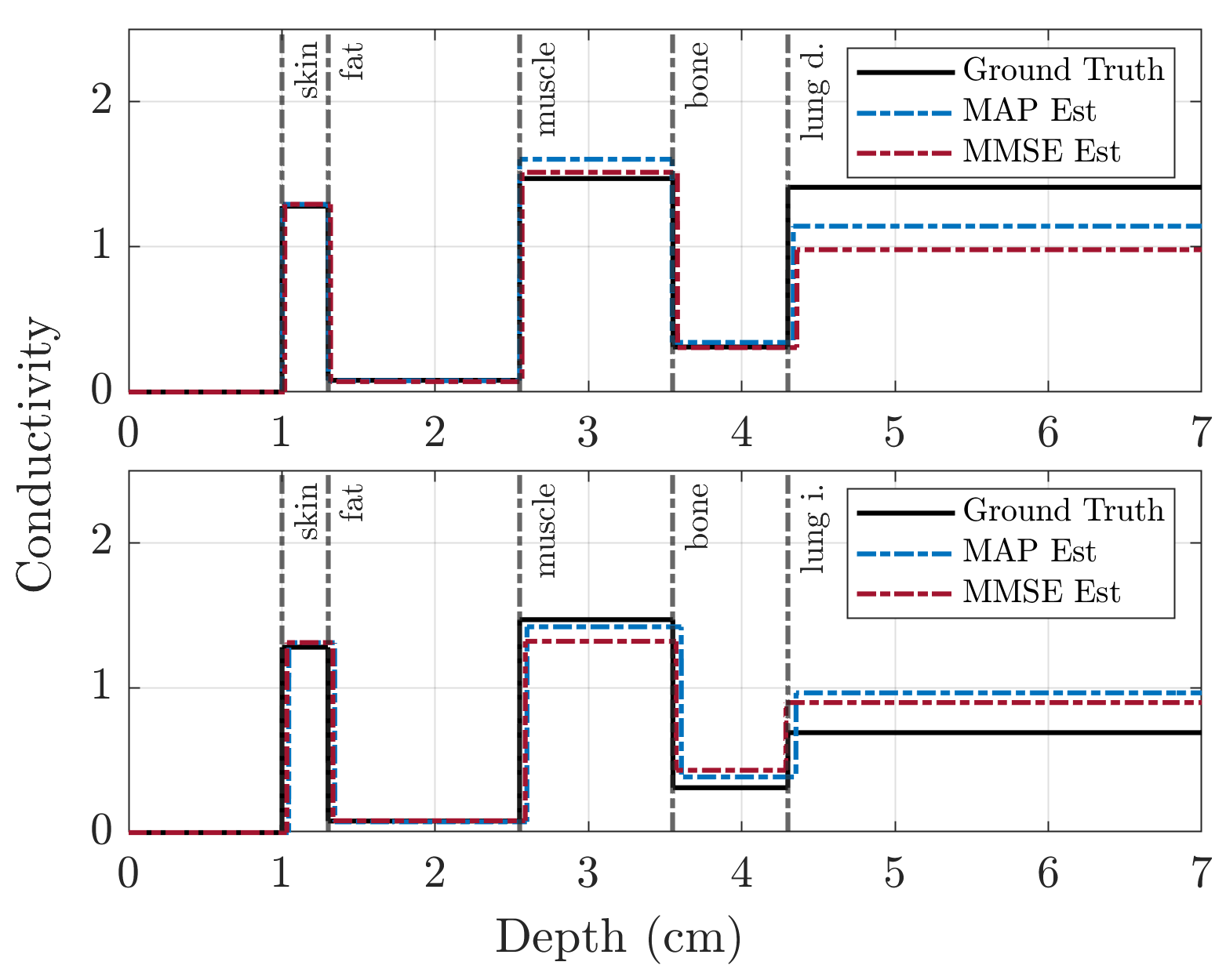}
  \end{subfigure}
  \begin{subfigure}[b]{0.32\textwidth}
  \centering
    \includegraphics[width=1\textwidth]{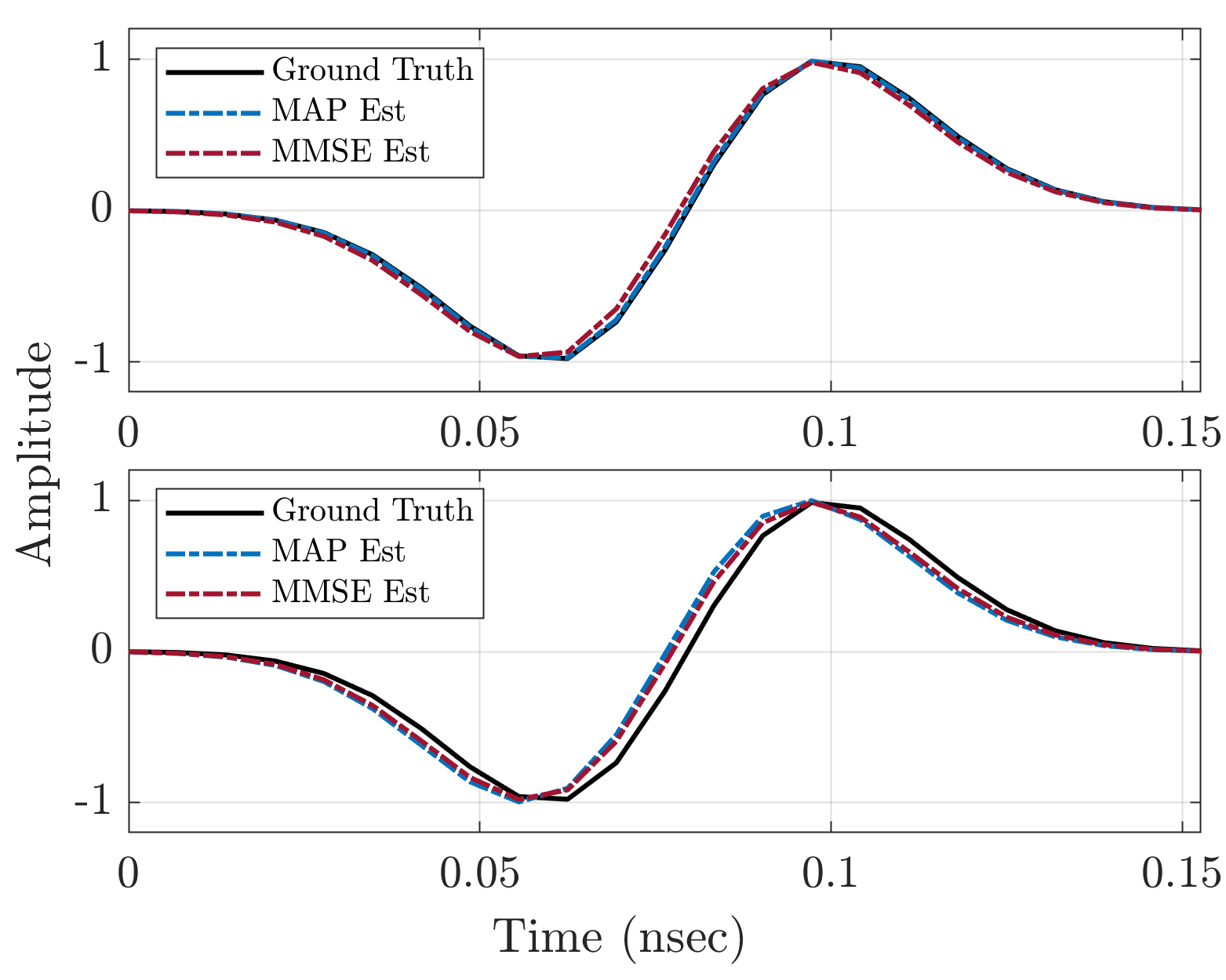}
  \end{subfigure}
  \caption{Recovery of relative permittivity profile (left), conductivity profile (middle), and transmitted pulse (right) for deflated (top) and inflated (bottom) lung scenarios at 40 dB SNR. \label{fig_example_recovery}}
  \vspace{-2mm}
\end{figure*}

\begin{figure*}[t!]
\centering
  \begin{subfigure}[b]{0.24\textwidth}
  \centering
    \includegraphics[width=\textwidth]{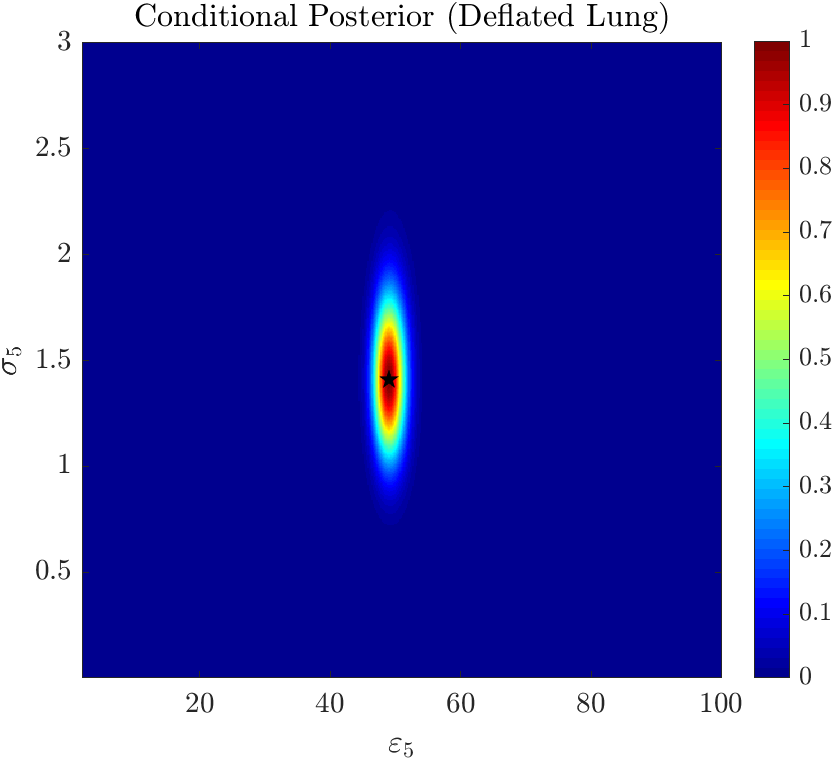}
  \end{subfigure}
  \begin{subfigure}[b]{0.24\textwidth}
  \centering
    \includegraphics[width=1\textwidth]{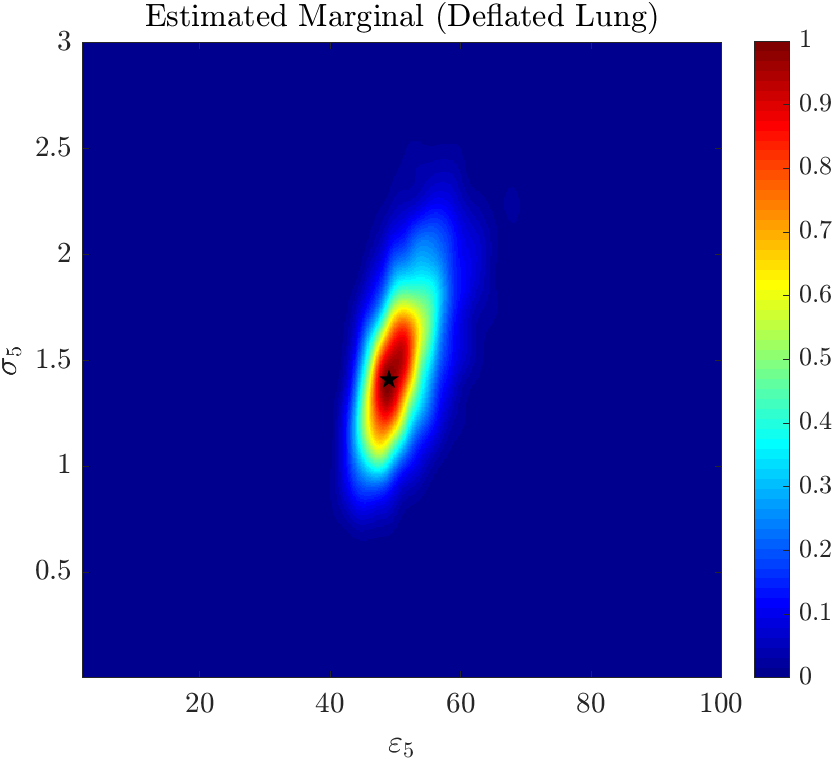}
  \end{subfigure}
  \begin{subfigure}[b]{0.24\textwidth}
  \centering
    \includegraphics[width=1\textwidth]{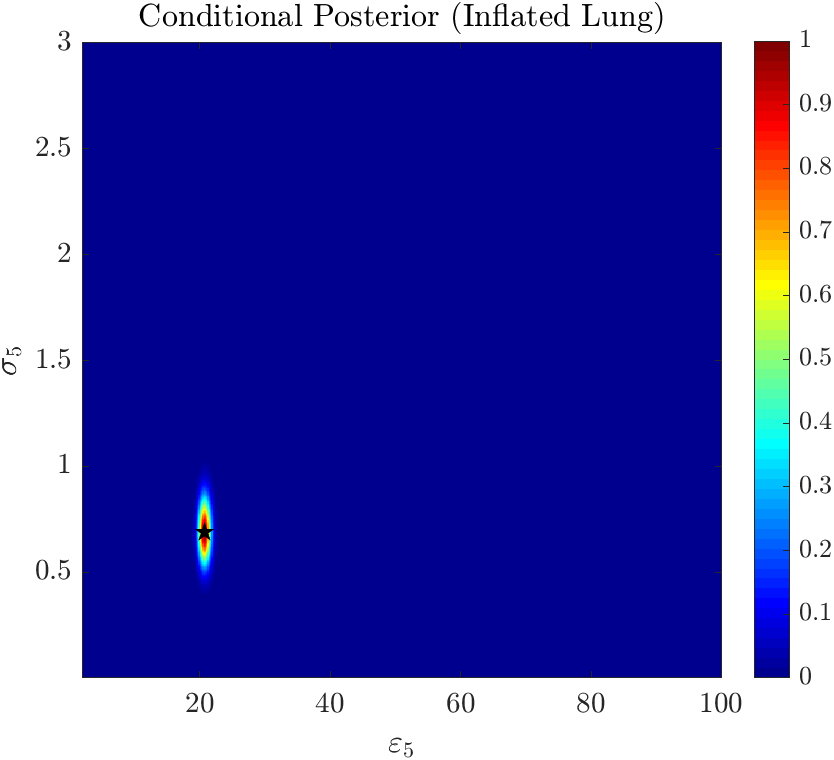}
  \end{subfigure}
  \begin{subfigure}[b]{0.24\textwidth}
  \centering
    \includegraphics[width=1\textwidth]{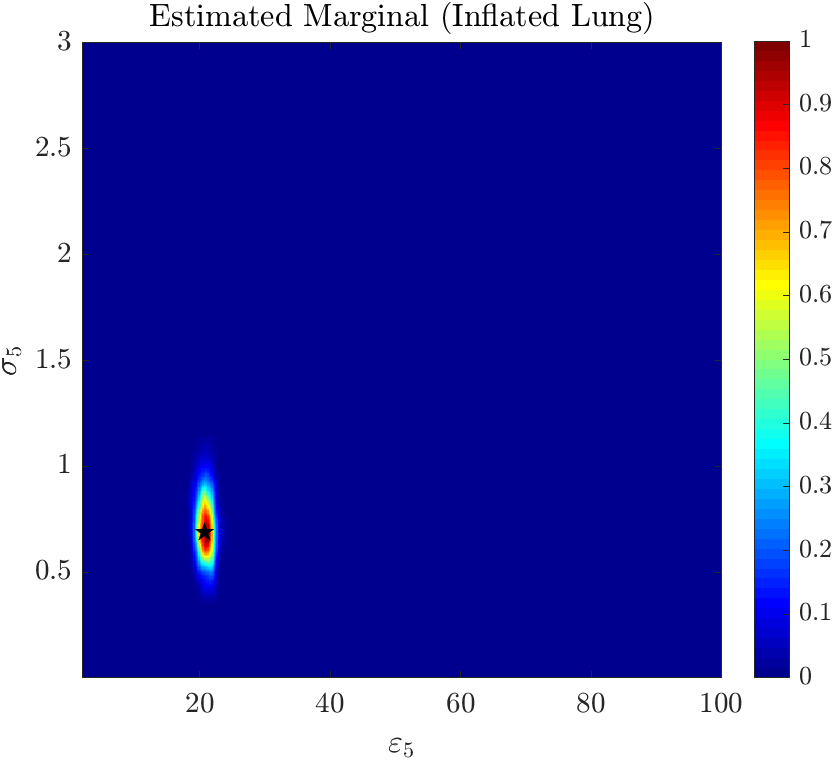}
  \end{subfigure}
    \caption{Actual conditional posterior densities and estimated marginal posterior densities of $\varepsilon_5$ and $\sigma_5$ for deflated and inflated lung scenarios. \label{fig_estimated_2D_densities}}
   \vspace{-3mm}
\end{figure*}

\subsection{Comparison of Sampling Efficiency}
After convergence to the stationary distribution, efficiency of a sampler can be measured based on the correlation of generated samples. In general, the consecutive samples generated within a MCMC scheme are correlated. Obviously, a lower correlation is more desirable since it increases the number of effective samples. It is defined as the ratio of total number of generated samples to the autocorrelation time (ACT). Therefore, ACT of a sampler provides an objective metric for comparing the efficiency of different samplers. It can be calculated by integrating the autocorrelation function (ACF), which is estimated over the chain of generated samples. The details of ACT and ACF calculations are provided in Section V of the supplemental material.\par

In Fig. \ref{fig_ACF_comparison}, we illustrate the estimated ACFs over the chains with $T_1 = 1$ for the first layer parameters $\varepsilon_1$, $\sigma_1$, and $d_1$. The ACFs were calculated on the converged portion of the chains which corresponds to Stage IV of our sampling scheme. The random walk behaviour of MH can be clearly observed by noting the existence of significant correlations even after long lags. Even though SS achieves a better sampling performance for $\sigma_1$ compared to HMC-\textbf{I}, they both perform very similarly for $\varepsilon_1$ and $d_1$, and still exhibit considerable correlations. On the other hand, HMC-$\vec{\hat{\Sigma}}_{SS}$ dramatically outperforms the others by rapidly vanishing the correlations. This indicates that employing a warm-up stage for covariance estimation considerably improves the sampling efficiency.  \par 

In order to have an analytical measure, we also compared the ACTs associated with each model parameter in Table \ref{tab_ACT_comparison}. As it can be observed, HMC-$\vec{\hat{\Sigma}}_{SS}$ dramatically reduces the number of samples required for generating a new independent sample for all model parameters. In addition, as the ACTs are highly fluctuating for different parameters in the case of other samplers, HMC-$\vec{\hat{\Sigma}}_{SS}$ provides a consistently lower ACT for all parameters. This is a natural result since the weighting matrix $\vec{M}$ successfully captures the linear correlations between different parameters. Overall, the obtained results demonstrate the superior sampling efficiency of HMC-$\vec{\hat{\Sigma}}_{SS}$.
\begin{figure*}[t!]
\centering
  \begin{subfigure}[b]{0.245\textwidth}
  \centering
    \includegraphics[width=1\textwidth]{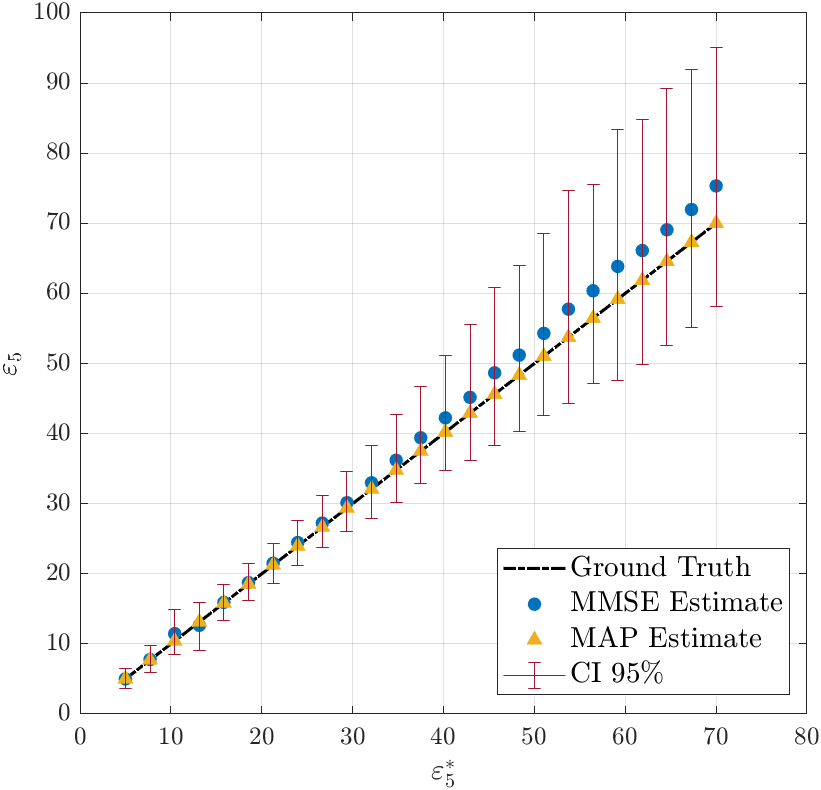}
  \end{subfigure}
  \begin{subfigure}[b]{0.245\textwidth}
  \centering
    \includegraphics[width=1\textwidth]{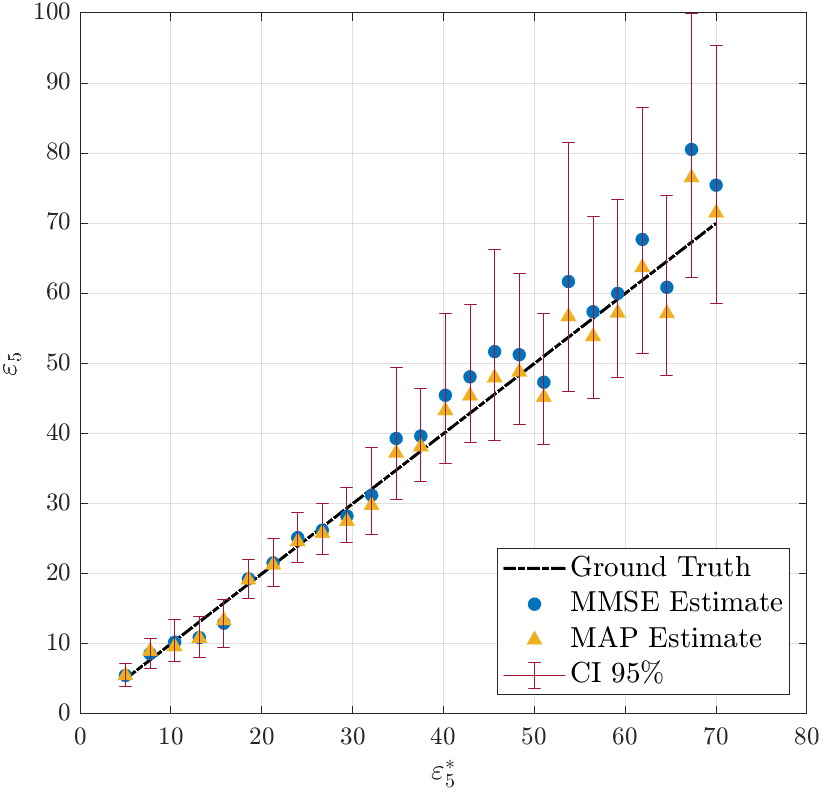}
  \end{subfigure}
  \begin{subfigure}[b]{0.245\textwidth}
  \centering
    \includegraphics[width=1\textwidth]{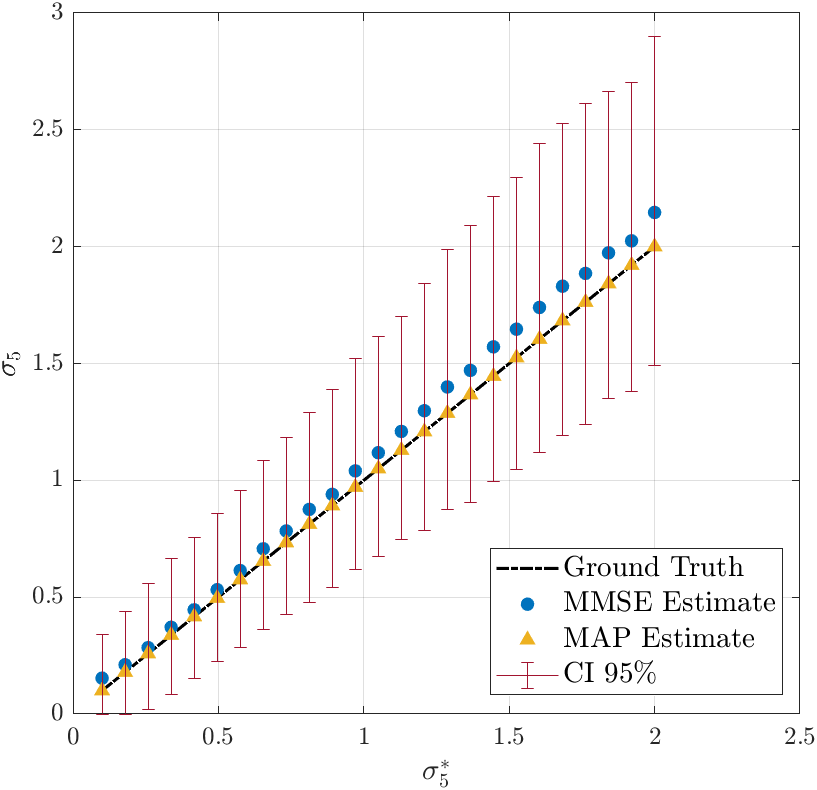}
  \end{subfigure}
  \begin{subfigure}[b]{0.245\textwidth}
  \centering
    \includegraphics[width=1\textwidth]{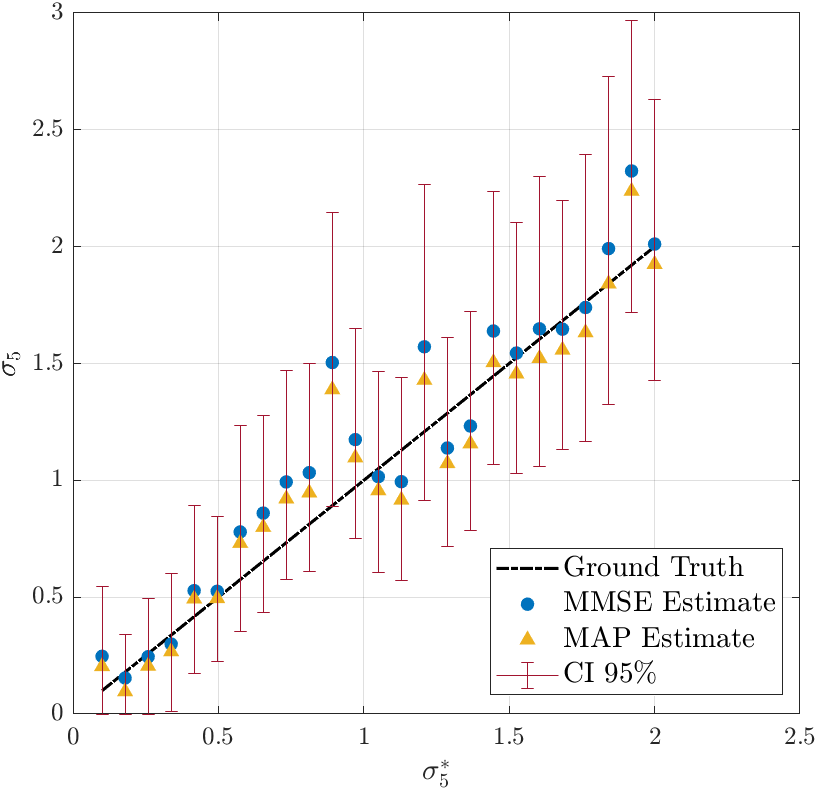}
  \end{subfigure}
  \caption{Estimation of last layer relative permittivity and conductivity along with $95\%$ credibility intervals corresponding to noise-free (first and third from left) and noisy (second and fourth from left) measurements at 40dB SNR. \label{fig_last_layer_param_tracking}}
\end{figure*}

\subsection{Validation of Self-Adaptation}
The adaptive models for temperature level and step size selection enable us to achieve improved sampling efficiency. To illustrate the adaptation process, in Fig. \ref{fig_adaptive_adjustment_schemes}, we represent the evolution of temperature levels and step sizes along with the associated swap and acceptance ratios. The lowest and highest temperature levels were fixed at $T_1 = 1$ and $T_{16} = 10^5$, and the remaining were initialized at geometrically spaced levels between $10^2$ and $10^3$ to better illustrate the evolution process. We initialize the sampling process with temperature adaptation using SS approach in Stage I. After the point at which the convergence criterion is satisfied, which is marked with the vertical dashed line located just after iteration 6000, we fixed the temperature levels and initiate Stage II. As it can be seen from the top left plot, the associated swap ratios between adjacent temperature levels successfully converge to a same level around 0.2. Once Stage II is finalized and an estimate of the covariance matrix is obtained, we initiate the step size adaptation with target acceptance ratio $\xi = 0.85$ for all temperature levels, as shown in the right plots of Fig. \ref{fig_adaptive_adjustment_schemes}. The step sizes were all initialized at $10^{-2}$, which is small enough to have roughly 100\% acceptance at each temperature level. As the evolution of acceptance ratios indicate, step sizes were successfully updated to achieve the desired acceptance ratio at all temperature levels until the convergence criterion is satisfied just before iteration 16000.\par 
We also provide example trace plots for parameters $\varepsilon_1$, $\sigma_1$, and $d_1$, corresponding each stage in Fig \ref{fig_sampling_illustration} to visually demonstrate the effect of adaptation stages on the sampling performance. During the first half of Stage I, we observe a strong random walk behaviour, especially for $\varepsilon_1$ and $d_1$, which is due to inadequate initialization of temperature levels. Once the temperatures are calibrated and the sampler converges to the stationary distribution, the random walk behaviour diminishes appreciably. But still, the generated sample traces exhibit noticeable correlations in Stage II, even though the sampling performance is visibly better compared to Stage I. In this stage, the sampling efficiency is limited by the performance of SS approach. After switching to HMC in Stage III, we again observe the random walk behaviour during the first a few hundreds of iterations due to inadequate selection of step sizes. However, as the step size adaptation progresses, HMC-$\vec{\hat{\Sigma}}_{SS}$ rapidly improves the sampling efficiency and starts producing samples with significantly reduced correlation. 

\subsection{Recovery Results on Synthetic Measurements}
In this part of the experiments, we assess the recovery performance of the proposed methods on synthetic measurements. The measurement sequences are created using the circular convolution model given in (\ref{measurement_model}). The reflectivity profiles are calculated using the 1D multilayer propagation model given in (\ref{multilayer_reflection_model}). We considered a multilayer structure with the following 5 layers:  skin (0.3 cm), fat (1.25 cm), muscle (1 cm), bone (0.75 cm), and lung (semi-infinite) to simulate human tissues in thoracic cavity. The actual typical permittivity and conductivity properties of each tissue were obtained from \cite{Gabriel_1996}. The transmitted waveform used in the experiments is the first derivative of Gaussian pulse with center frequency $f_c = 4$ GHz, which is nearly bandlimited with a bandwidth of $4$ GHz.\par

As an illustrative example, in Fig. \ref{fig_example_recovery}, we represent the recovery results for the relative permittivity and conductivity profiles as well as the transmitted waveform using the measurement with $40$ dB SNR. We note that such high level of SNR is required for observing meaningful reflections from deeper tissues. We included both deflated and inflated lung scenarios to investigate whether it is possible to detect variations in the last layer parameters. We used the sample mean of the generated samples as an approximation to the MMSE estimate, while we employed gradient based off-the-shelf local search methods initialized at the sample that achieves the highest posterior value to for the MAP estimate. The recovered profiles indicate that estimating relative permittivity is relatively easier as opposed to estimating conductivity property. Moreover, the thickness estimation is almost perfect for all layers. This is mainly due to the shape of posterior distribution. In order to justify this, we illustrate the true conditional 2D posterior distributions of $\varepsilon_5$ and $\sigma_5$, where all other parameters are fixed at their true values, as well as the corresponding estimated 2D marginal distributions in Fig. \ref{fig_estimated_2D_densities}. The results points out that the variance along $\sigma_5$ direction is considerably higher, making successful recovery more difficult. Nevertheless, we also observed from conditional distributions that the modes of posterior distribution are clearly separated for deflated and inflated lung scenarios, which is successfully captured by the estimated marginal distributions as well. This indicates the possibility of detecting variations in deeper tissue layers given sufficiently high SNR in blind setting, where the transmitted waveform is almost perfectly recovered in both cases as well. As a final note, we did not observe a remarkable difference between MMSE and MAP estimates, which can be explained by the nearly symmetric structure of the estimated marginals. 

\begin{figure*}[t!]
\centering
  \begin{subfigure}[b]{0.32\textwidth}
  \centering
    \includegraphics[width=0.98\textwidth]{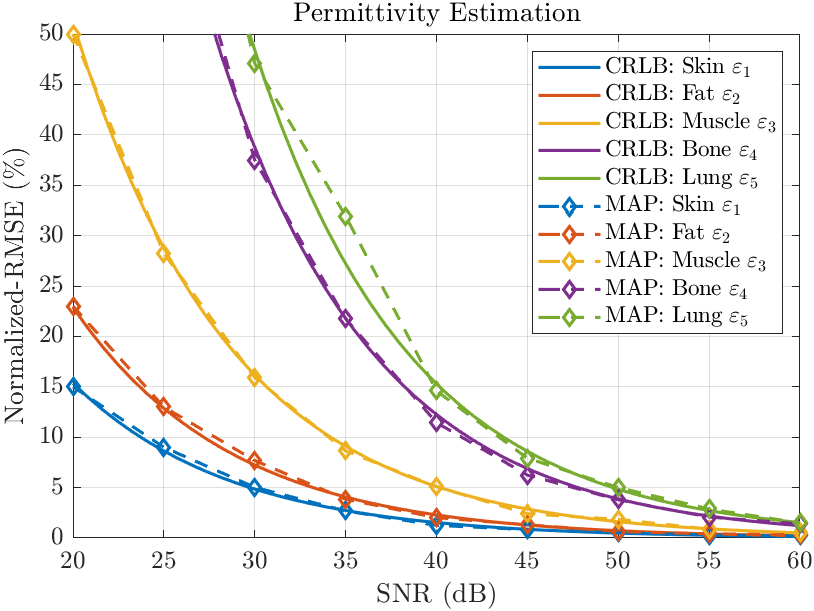}
  \end{subfigure}
  \begin{subfigure}[b]{0.32\textwidth}
   \centering
    \includegraphics[width=0.98\textwidth]{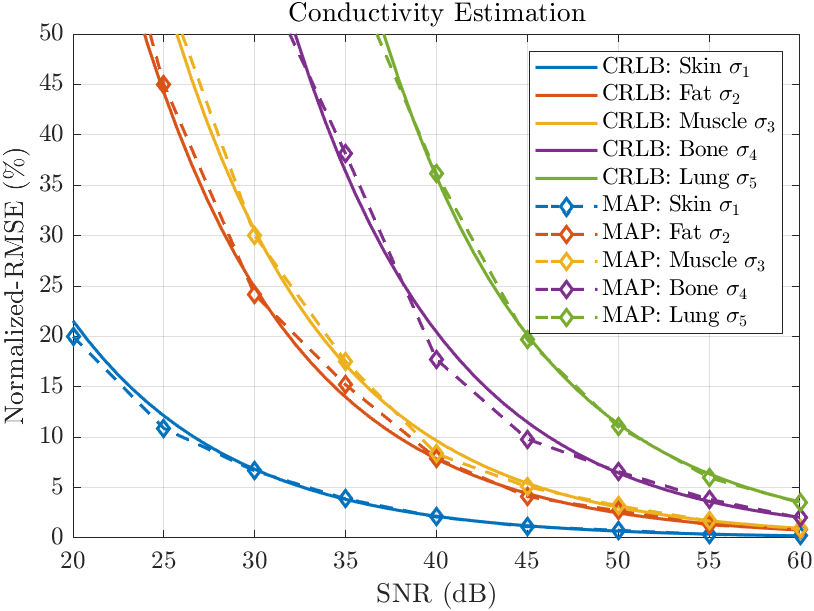}
  \end{subfigure}
  \begin{subfigure}[b]{0.32\textwidth}
   \centering
    \includegraphics[width=0.98\textwidth]{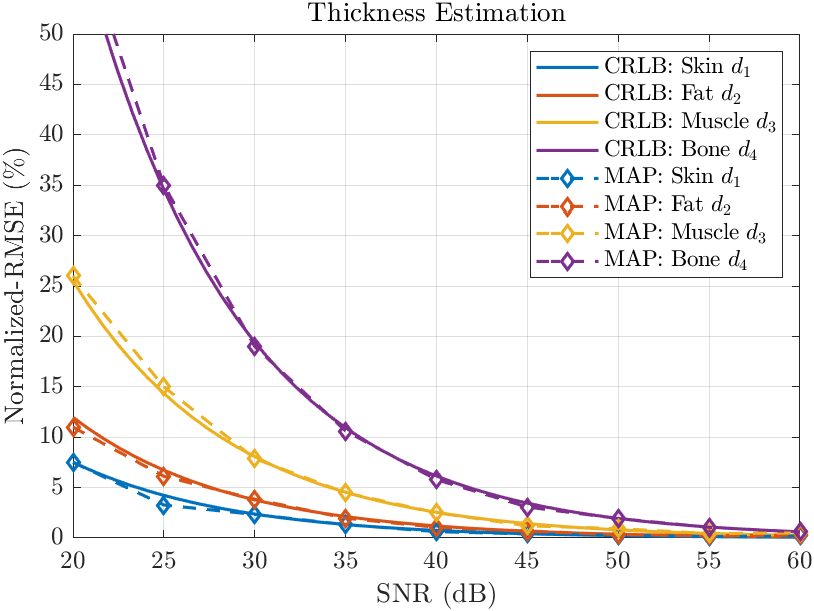}
  \end{subfigure}
  \begin{subfigure}[b]{0.32\textwidth}
  \centering
    \includegraphics[width=0.95\textwidth]{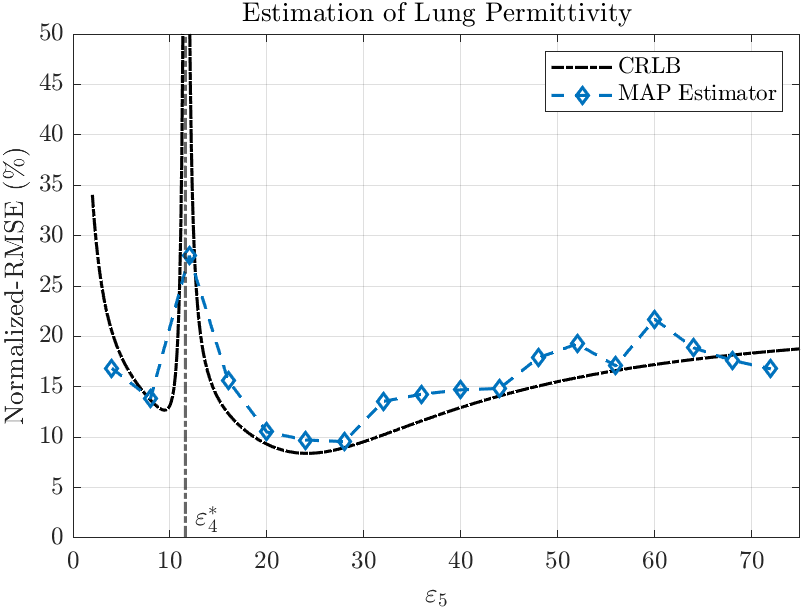}
  \end{subfigure}
  \begin{subfigure}[b]{0.32\textwidth}
  \centering
    \includegraphics[width=0.95\textwidth]{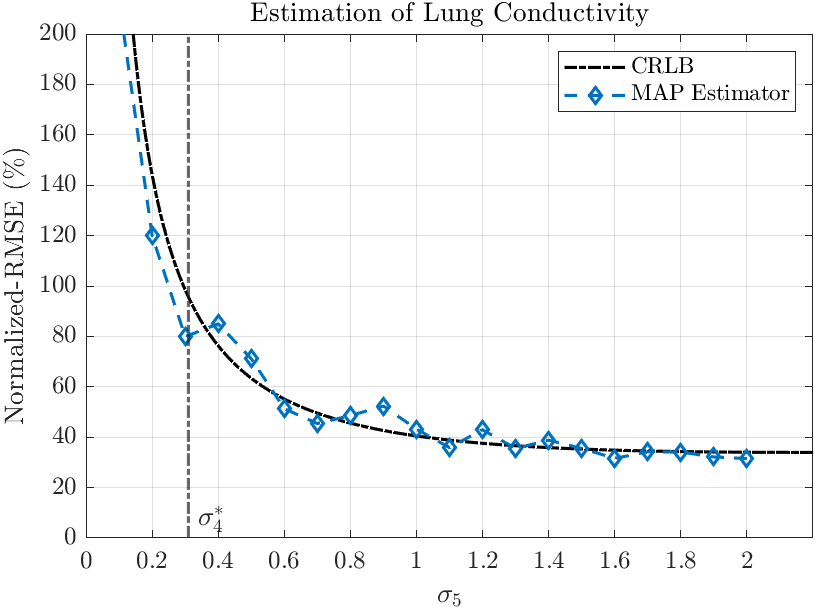}
  \end{subfigure}
  \caption{Comparison of the CRLB and the MAP estimator variance. Flat priors were used to mimic ML estimator. The estimator variance is empirically calculated based on 100 noisy observations. Top figures illustrate the Normalized RMSEs as a function of SNR for all model parameters. Bottom figures represent the Normalized RMSEs as a function of the last layer parameter values at 40 dB SNR. \label{fig_CRLB_vs_MAP_variance}}
\end{figure*}

\subsection{Estimation with Credibility Intervals}
We now consider two different scenarios, where in the first one, we varied the last layer relative permittivity in between 5 and 70, and in the second one, we varied the last layer conductivity in between 0.125 and 2, while keeping all other parameters constant at their typical values. Our goal is to investigate the tracking performance of the estimators. For these experiments, on top of the point estimates of MMSE and MAP, we also compute the 95\% credibility intervals to represent the uncertainty of estimates. We considered both noise-free and noisy (40 dB SNR) measurement cases to see how the estimates and the associated credibility intervals change. The noise-free measurement was still handled within the noisy model, i.e., it just represent the \textit{lucky} case, where the noise components were happened to be zero at all indices. \par 

We demonstrate the recovery results along with the credibility intervals in Fig. \ref{fig_last_layer_param_tracking}. Considering the noise-free scenarios, the MAP estimate perfectly recovers the actual parameter values. This is an expected result since the prior distributions were selected in a way not to disturb the mode of the posterior distribution. For noisy measurements, the MAP estimates fluctuate around the true values due to disturbed mode of the posterior. The MMSE estimates seem to be consistently overestimating in all cases, especially for larger values of $\varepsilon_5$ and $\sigma_5$. This is an indicator that posterior distributions are skewed towards larger parameter values. Hence, the MAP estimator might be a more favorable choice over MMSE. The credibility intervals provide useful information about the shape of distributions. It can be observed that the posterior becomes more peaky around the true values for smaller values of $\varepsilon_5$ and $\sigma_5$, which was also observed in Fig. \ref{fig_estimated_2D_densities}. Hence, one might argue that it is relatively easier to estimate smaller values of parameters, especially for relative permittivity. Our final observation is, for noisy measurements, the results show that actual parameter value always lie within the 95\% credibility interval at 40 dB SNR. 

\subsection{Theoretical Bounds on the Estimator Performance}
In order to assess the estimation performance, in this section, we derive the Cramer-Rao Lower Bounds (CRLB) for unbiased estimators and present the best achievable performance on estimation of tissue properties in blind setting. We assume that non-informative flat prior distributions are employed for the multilayer model parameters (with $\kappa_i = 0$) and that the variance of pulse subspace $\sigma_{\gamma}^2$ is sufficiently high. In this setting, the problem can be considered within the frequentist approach and the unknown parameters can be treated as deterministic valued quantities. Let us collect all the parameters except the noise variance in $\vec{\phi} = (\vec{\theta},\vec{\gamma})$ and denote the noise-free signal as $\vec{s} = \text{diag}(\vec{F}_Q\vec{h})\vec{x}$, which is then corrupted by white Gaussian noise $\vec{v}$. For a given noise variance $\sigma_v^2$, the log-likelihood is expressed as
\begin{equation}\label{crlb_derivation}
    \log p(\vec{y}|\vec{\phi}) = -N\log(\pi\sigma_v^2)-\dfrac{1}{\sigma_v^2}\sum_{n=0}^{N-1}|y_n - s_n|^2,
\end{equation}
where the partial second derivatives are given by 
\begin{equation}
    \dfrac{\partial^2 \log p(\vec{y}|\vec{\phi})}{\partial \phi_i \partial \phi_j} = \dfrac{2}{\sigma_v^2}\sum_{n=0}^{N-1}\Re\bigg\{(y_n - s_n)^*\dfrac{\partial^2 s_n}{\partial \phi_i \partial \phi_j} -\dfrac{\partial s_n^*}{\partial \phi_j}\dfrac{\partial s_n}{\partial \phi_i}\bigg\}.
\end{equation}
For multivariate case, the Fisher information matrix $\mathcal{I}(\vec{\phi})$ has the following form 
\begin{equation}\label{fim}
    [\mathcal{I}(\vec{\phi})]_{i,j} = -E\bigg[\dfrac{\partial^2 \log p(\vec{y}|\vec{\phi})}{\partial \phi_i \partial \phi_j}\bigg] = \dfrac{2}{\sigma_v^2}\sum_{n=0}^{N-1}\Re\bigg\{\dfrac{\partial s_n^*}{\partial \phi_j}\dfrac{\partial s_n}{\partial \phi_i}\bigg\},
\end{equation}
since $E[y_n] = s_n$. Here, $[\cdot]_{i,j}$ denotes the element at $i^{th}$ row and $j^{th}$ column. Therefore, the covariance matrix $\vec{C}_{\hat{\phi}}$ of any unbiased estimator $\hat{\vec{\phi}}(\vec{y})$ satisfies $\vec{C}_{\hat{\phi}} - \mathcal{I}^{-1}(\vec{\phi}) \succcurlyeq 0$, i.e.,
\begin{equation}
    \text{Var}(\hat{\phi}_i) = [\vec{C}_{\hat{\phi}}]_{i,i} \geq [\mathcal{I}^{-1}(\vec{\phi})]_{i,i}.
\end{equation}
The derivations for partial derivatives in (\ref{fim}) are provided in Section IV of the supplementary material. Although the recursive structure of the derivatives prevents obtaining analytical expressions, we can still calculate the CRLBs numerically.\par

In upper plots of Fig. \ref{fig_CRLB_vs_MAP_variance}, we present the minimum achievable Normalized Root Mean Square Error (N-RMSE) as a function of SNR for each of the multilayer model parameters. We also included the empirically estimated N-RMSE of our MAP estimator, which uses flat priors to mimic the Maximum-Likelihood (ML) estimator. The empirical error rates were estimated over 100 different noisy measurements generated with the same model parameters. The first and most essential observation is that the MAP estimator strictly achieves the CRLB for the given range of SNRs for all parameters. Secondly, the error rates are consistently higher for deeper tissues, which is an expected result due to considerable signal attenuation. Comparing the estimation of different sets of parameters, we observe that the lowest achievable error rates are for thicknesses, followed by relative permittivities, and conductivities. Therefore, the posterior is much more sensitive to changes in the layer thicknesses as opposed to other properties. With this results, we are also able to quantify the expected recovery performance. For example, even with 40 dB SNR, the minimum achievable N-RMSE is around 15\% for lung permittivity and 36\% for lung conductivity. \par

In lower plots of Fig. \ref{fig_CRLB_vs_MAP_variance}, we presented the lower bounds as well as the empirical error rates of MAP estimator for different values of $\varepsilon_5$ and $\sigma_5$ at 40 dB SNR. The results show that the MAP estimator achieves the lower bounds even for different parameter values. One important observation is that when we have $\varepsilon_5 \approx \varepsilon_4$, the CRLB for $\varepsilon_5$ increases significantly. The main reason for this phenomenon can be explained as follows. When the relative permittivities of adjacent layers are indistinguishably close, the magnitude of the reflection coefficient at that interface becomes considerably small, and hence, the actual 5-layer model behaves like a 4-layer structure, causing overparametrization. This result indirectly informs us about the recovery performance when using more number of layers than the underlying model itself has. Unlike the relative permittivity, we do not observe the same phenomenon in the case of conductivities, which is most likely due to the fact that conductivity difference has a minor effect on the magnitude of reflection coefficients. 

\section{Concluding Remarks}\label{sec_conclusion}
In this paper, we studied the reconstruction of one-dimensional multilayer tissue profiles from ultrawideband radar measurements. We assumed a blind setting and jointly estimated both the transmitted radar waveform and the multilayer model parameters. We approached the problem from a Bayesian perspective and presented a comprehensive MCMC method to perform inference on the highly complex posterior distribution. We employed parallel tempering to resolve the local optimality issue, estimated covariance of the posterior to capture linear correlations between model parameters, and incorporated adaptation methods to adjust the sampler parameters. As a result, the proposed sampling mechanism achieved superior sampling efficiency compared to conventional sampling schemes. Simulations on the synthetic radar measurements revealed successful recovery results. Comparisons with the derived theoretical bounds showed that the proposed estimator achieves the minimum possible error rate. More importantly, the estimated marginal posterior distributions revealed promising results indicating the feasibility of tracking/detecting variations in deeper tissue layers. Overall, although the one-dimensional setting investigated in this work is a simplified version of the reality, it provides useful insights about the feasibility and challenges of the problem. As the future work, we aim to extend the presented recovery methods to a three-dimensional wave propagation model, which has been rigorously studied in \cite{LambotS,LambotS1,LambotS2} for ultrawideband radar systems. In addition, we also aim to incorporate frequency dependence of model parameters through Debye relaxation models \cite{DebyeP} to improve modelling accuracy.

\bibliographystyle{IEEEtran} 
\bibliography{References}

\end{document}